\documentclass[twocolumn,epsfig,aps]{revtex4}   
\usepackage{epsfig}
\usepackage{graphicx}

\begin{document}
\title{Volume-Enclosing Surface Extraction}
\author{B.R.~Schlei\footnote{\textit{Email address:} \texttt{schlei@me.com}}}
\affiliation{
GSI Helmholtz Centre for Heavy Ion Research GmbH,
Planckstra\ss e 1, 64291 Darmstadt, Germany
}
              
\begin{abstract}
In this paper we present a new method, which allows for the construction of 
triangular isosurfaces from three-dimensional data sets, such as 3D image data
and/or numerical simulation data that are based on regularly shaped, cubic 
lattices.
This novel volume-enclosing surface extraction technique, which has been named 
VESTA, can produce up to six different results due to the nature of the discretized
3D space under consideration.
VESTA is neither template-based nor it is necessarily required to operate on
$2\times2\times2$ voxel cell neighborhoods only. 
The surface tiles are determined with a very fast and robust construction
technique while potential ambiguities are detected and resolved. 
Here, we provide an in-depth comparison between VESTA and various versions of
the well-known and very popular Marching Cubes algorithm for the very first time. 
In an application section, we demonstrate the extraction of VESTA isosurfaces
for various data sets ranging from computer tomographic scan data to simulation
data of relativistic hydrodynamic fireball expansions.\\


\end{abstract}
\maketitle

\section{Introduction}

\noindent
The determination of implicit surfaces, which are contained in three-dimensional
(3D) image data and numerical 3D simulation data that are
based on regularly shaped, cubic lattices, has become an important tool within 
many fields of science, industry and medicine (\textit{cf.}, e.g., 
Ref.s~\cite{RUSS98, LOHM98}).
Such (usually triangular) surfaces can be used for visualization 
purposes~\cite{FOLE93}, e.g., when 3D shapes that are contained in the data 
should be rendered, and/or they may represent the basis for further numerical 
evaluations.
To this date, many articles have been written on the generation of triangular
surfaces from 3D digital data sets (\textit{cf.}, e.g., Ref.s~\cite{LOHM98,
LORE87, NVID07, DOI91, GUEZ95, THIR96, BLOO94, HO05, HILL95, CHER95, LEWI03,
NEWM06}).\\
\indent
A very popular algorithm for surface construction from 3D image data has been
provided through the Marching Cubes algorithm (MCA), which has been developed by
Lorensen and Cline in the mid-1980s~\cite{LORE87}.
Note, that this tool is nowadays still advertised (e.g., by the multinational 
NVIDIA Corporation~\cite{NVID07}) as one of the state-of-the-art technologies 
for digital surface construction.
Alternate approaches for 3D surface construction include -- but are not limited 
to -- the Marching Tetrahedrons~\cite{DOI91, GUEZ95} (MTA),
Marching Lines~\cite{THIR96} (MLA; \textit{cf.}, also Ref.~\cite{BLOO94}),
the Cubical Marching Squares~\cite{HO05} (CMSA) algorithms, and
VESTA~\cite{BRS03, VEST04}, which is described in this paper.\\
\indent
The MCA is a template-based approach, and as a consequence of the non-trivial
topology in 3D~\cite{HILL95}, several surface templates have been initially 
overlooked, resulting in the accidental generation of holes in some data 
sets~\cite{LOHM98}.
Meanwhile, this problem has been fixed (\textit{cf.}, e.g., 
Ref.s~\cite{CHER95,LEWI03}).
For a recent and very detailed discussion on the MCA's history, \textit{cf.},
Ref.~\cite{NEWM06}.
The MTA, unfortunately, has directional ambiguities, because it subdivides a cube
with tetrahedrons, which cannot be done symmetrically~\cite{DOI91, GUEZ95}.
The MLA -- and the similar techniques -- are not template 
based~\cite{THIR96, BLOO94}, but they require further processing, so that
apparent cell ambiguities are properly resolved (\textit{cf.}, e.g.,
Ref.~\cite{HILL95}).
The CMSA is very similar to the previous, cubically based approaches, however
templates are used for the faces of the $2\times2\times2$ voxel neighborhoods
under consideration, and further work to resolve cell ambiguities is done as 
well~\cite{HO05}.\\
\indent
The "Volume-Enclosing Surface exTraction Algorithm" (VESTA), which is
presented in this paper, allows one to numerically compute -- very fast and
totally robust (i.e., without the accidental generation of any holes) -- 
non-degenerate, mathematically well oriented triangular surfaces in 3D~\cite{BRS03}.
VESTA constructs surfaces -- so to speak -- from the grounds up, because it
continuously transforms the initial surface that consists of the boundary faces 
(squares) of selected voxels into the final isosurface.
Instead of using many different surface templates, VESTA uses a single building 
block that is based on the vector decomposition of a single voxel face 
(\textit{cf.}, Fig.~2.b).
VESTA collects all of the participating vectors and groups them into closed 
vector cycles, while resolving 3D cell ambiguities properly.
In fact, this is done in analogy to the DICONEX algorithm~\cite{BRS09}, 
which allows one to construct gap-free contours for 2D digital data.\\
\indent
The original VESTA algorithm is not limited to the processing of 
$2\times2\times2$ voxel cell neighborhoods, and it therefore minimizes
the generation of redundant information.
Since VESTA surfaces can be generated in a global ``disconnect'', a global
``connect'', and a ``mixed'' connectivity mode, and since all of these
three modes can be executed either in a low resolution (``L'') or in a high 
resolution (``H'') mode, VESTA can produce in total six different types of 
surfaces on demand.
Since all VESTA surface cycles are confined to $2\times2\times2$ voxel cell 
neighborhoods at all times (\textit{cf.}, below), a marching $3$-cell 
variant is presented in this paper as well.
The marching VESTA makes use of a very simple, but explicit table (\textit{cf.}, Table~1, subsection II.F.) of directed vector paths for the 
construction of closed vector cycles,
and it is therefore very easy to implement into computer code.\\
\indent
This paper is organized as follows.
First, we shortly review the DICONEX algorithm~\cite{BRS09}, which uses a 2D 
digital data set as input, because VESTA can be viewed as an extension of 
DICONEX from 2D into 3D.
Then we shall explain VESTA in great detail.
In particular, we shall compare VESTA with both: 
(i) the original MCA~\cite{LORE87}, which can be found in many text books 
(\textit{cf.}, Ref.s~\cite{LOHM98,NVID07}), but which may generate holes; 
and
(ii) a state-of-the-art MCA implementation~\cite{BOUR94} by Bourke et al., which 
represents an extended version of the original MCA that does \textit{not} 
generate any holes and that produces exactly the same surface as VESTA, when it 
is executed in its low resolution (``L''), global ``disconnect'' mode.
The in-depth comparison with the original MCA is done within the theoretical
section, whereas the comparison with the extended MCA is done within
three VESTA application related subsections. 
Finally, this paper will conclude with a short summary.

\section{The Surface Extraction Framework}

\noindent
Before we explain the surface extraction with VESTA, we first consider contour
extraction in 2D with DICONEX.

\begin{figure}[b]
\epsfig{width=8.5cm,figure=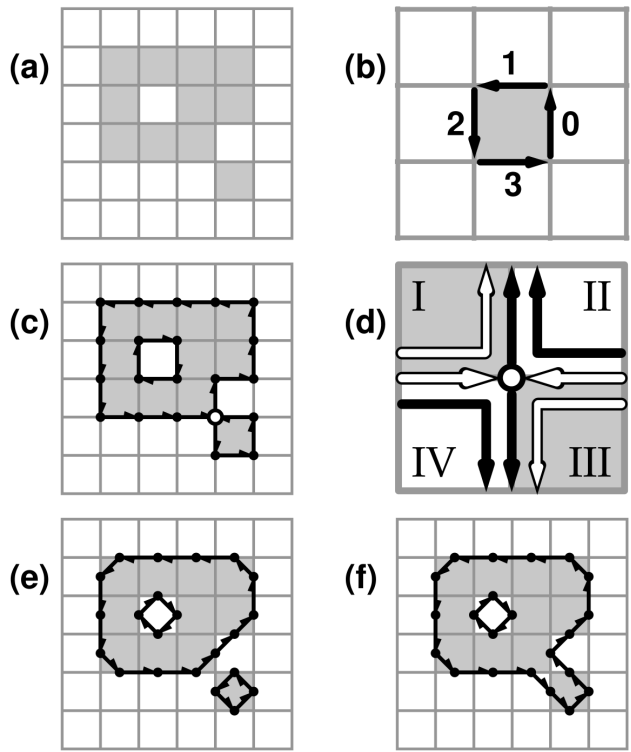}
\vspace{-0.1in} 
\caption{
(a) Binary image; 
(b) initial contour vectors for a single pixel; 
(c) initial contour vectors and junctures (black dots) for the binary image;
(d) connection diagram for a single point of ambiguity (white dot);
(e) dilated contour vectors and support points (black dots) resulting from the
``disconnect'' mode; 
(f) as in (e), but resulting from the ``connect'' mode.
} \label{multib_01}
\end{figure}

\subsection{DICONEX - {\underline DI}lated {\underline CON}tour 
{\underline EX}traction}

\noindent
In Fig.~1.a, we show a binary image with $6\times6 = 36$ pixels (i.e., 
picture elements).
Let us assume, that the gray pixels have been segmented, i.e., they have
been selected for an enclosure.
In the following, we shall denote segmented pixels as ``active'' and the
other pixels as ``inactive'' pixels.
The objective is now to enclose the $11$ active (gray) pixels with contours
while making use of the DICONEX algorithm~\cite{BRS09}.
First all initial contour vectors (ICV) ought to be collected.
The ICVs are here oriented pixel edges that separate an active pixel from 
an inactive one.\\
\indent
E.g., for a single active pixel that has no active next neighbor at all
(\textit{cf.}, Fig.~1.b), we have a maximum of four single ICVs (in the figure,
they are numbered from $0$ to $3$). 
Note, that an active pixel lies always to the left of its corresponding ICV.
Because of this construction, it is possible to forward the information of the
interior and/or exterior of a shape (i.e., a collection of segmented pixels) 
that needs to be enclosed by a finite number of contours.
Note that a shape is enclosed counterclockwise, while its potential holes are
enclosed clockwise.\\
\indent
In a second step, the ICVs will be connected to oriented shape-enclosing
contours.
Note, that each ICV both starts and ends in a single juncture (\textit{cf.},
Fig.~1.c). 
Most of the times it is trivial to find the successor of an ICV, if we
attempt to determine the particular IVC connectivities. 
But sometimes it is possible that a single active pixel is in contact with
another active pixel through only one single point (\textit{cf.},
white dot in Fig.~1.c). 
The latter results in a situation where we have two incoming and 
two outgoing ICVs for this particular point of contact (\textit{cf.},
Figs.~1.c and~1.d).
In the following, we shall call such a juncture ``point of ambiguity'' (POA). 
In order to avoid gaps in the final set of contours, one has to treat  
these POAs in a special way (\textit{cf.}, Ref.~\cite{BRS09}).\\
\indent
In Fig.~1.d, a connectivity diagram is depicted. 
If one connects to an incoming IVC an outgoing IVC that belongs to the same
active pixel as for the incoming IVC, then one performs a left turn, i.e.,
one follows one of the paths of the white bent vectors.
Conversely, if one connects to an incoming IVC an outgoing IVC that belongs to
another active pixel as for the incoming IVC, then one performs a right turn, 
i.e., one follows one of the paths of the black bent vectors instead.
The particular turns will either lead to a separation (``disconnect'' mode), 
or to a joining (``connect'' mode) of shapes next to a POA, respectively.\\
\indent
A user can interactively make this consistent choice of either left or
right turns within one given image, e.g., if only binary information is
available, in order to ensure the construction of {\it gap-free} contours.
Note that this rather global decision making process may be replaced and
automated by a local decision on the turning of the IVCs, while using, e.g.,
the average gray level of the four pixels, which surround a particular POA,
provided that gray level information is available.
In Fig.s~1.e and~1.f, the final DICONEX contours are depicted for
the ``disconnect'' mode and for the ``connect'' mode, respectively.
These contours have been obtained from a displacement of the origins of the
IVCs to the middle points of their corresponding pixel edges.\\
\indent
In essence, the DICONEX contours result from a displacement of the
initial pixel edges that separate active pixels from inactive ones.
Note that the set of these initial pixel edges already provides a
perfect enclosure of the segmented pixels.
In 3D, we shall proceed now analogously. 
 
\subsection{Initial Considerations for Voxels}

\noindent
VESTA will enclose voxels that have an inherent field value, e.g., a shade 
of gray, above an initially given threshold. 
A voxel (i.e., VOlume piXEL) is a 3D object.
To be more specific, it is represented by a cube and, hence, it has six
squares as faces. 
For an active voxel, i.e., a voxel that should be enclosed by a surface
section, VESTA will check all of its six nearest neighbors.
If there is a transition from an active to an inactive voxel, the
corresponding voxel face will be recorded.\\
\indent
In Fig.~2.a, two neighboring voxels are shown.
One voxel, i.e., the active voxel, is marked with a sphere at its center.
The second voxel has no sphere at its center, because it is considered
inactive, i.e., this voxel has a field value below the initially given
threshold. 
Between an active and an inactive voxel we shall have a contribution to
the enclosing surface.
Therefore, the face that separates these two voxels has to be considered.\\
\indent
In Fig.~2.b, we show such a boundary face.
The center of this face is marked with a black dot.
Such voxel face centers will finally be support points of the VESTA surface.
In case an isosurface should be constructed, the voxel face centers may
be moved within the bounds of its corresponding range vector, $r$,
which has its origin at the center of the active voxel and which ends
in the center of the neighboring inactive voxel.
Fig.~2.c helps to illustrate, how VESTA surfaces can be
transformed into isosurfaces while considering the 2D analog of
two neighboring pixels.

\begin{figure}[t]
\epsfig{width=8.5cm,figure=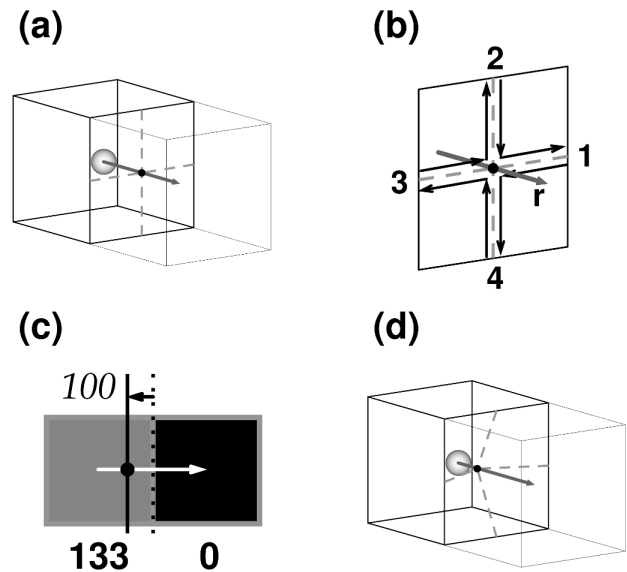}
\vspace{-0.1in} 
\caption{
(a) Two voxels (one is active (sphere) and one is inactive), which are in
contact through a voxel face, and a corresponding range vector;
(b) a voxel face with its voxel face center (black dot), its range 
vector, $r$, (gray), four edge middle points ($1, 2, 3, 4$, which each are 
connected with a dashed line to the voxel face center), and voxel face 
vectors (black);
(c) contour displacement for a pixel pair (see text);
(d) as in (a), but with a displaced voxel face center.
} \label{multib_02}
\end{figure}
\begin{figure*}[t]
\epsfig{width=15.0cm,figure=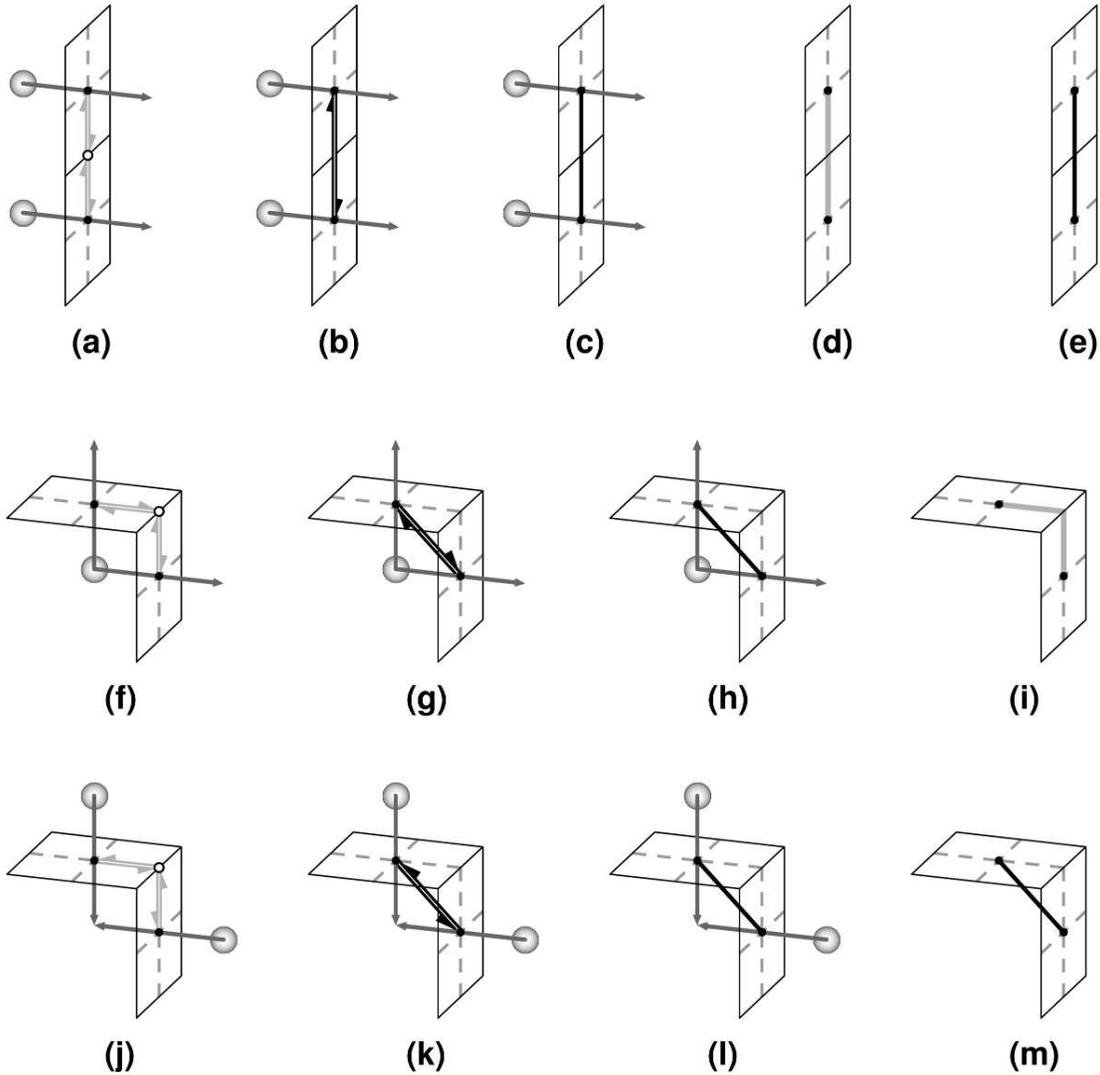}
\vspace{-0.1in} 
\caption{
Pairs of boundary faces, which are in direct contact (see text);
(a) -- (e) in-plane, for two active voxels each;
(f) -- (m) out-of-plane, for either one or three active voxels each;
the spheres indicate active voxels, which are shown only for
corresponding boundary faces, and in conjunction with their corresponding 
range vectors; 
light gray vector pairs and thick light gray lines represent voxel face 
vectors, whereas black vector pairs and thick black lines represent
VESTA cycle vectors;
black dots are VESTA surface support points, whereas white dots 
represent junctures. 
} \label{multib_03}
\end{figure*}
\begin{figure}[t]
\epsfig{width=8.0cm,figure=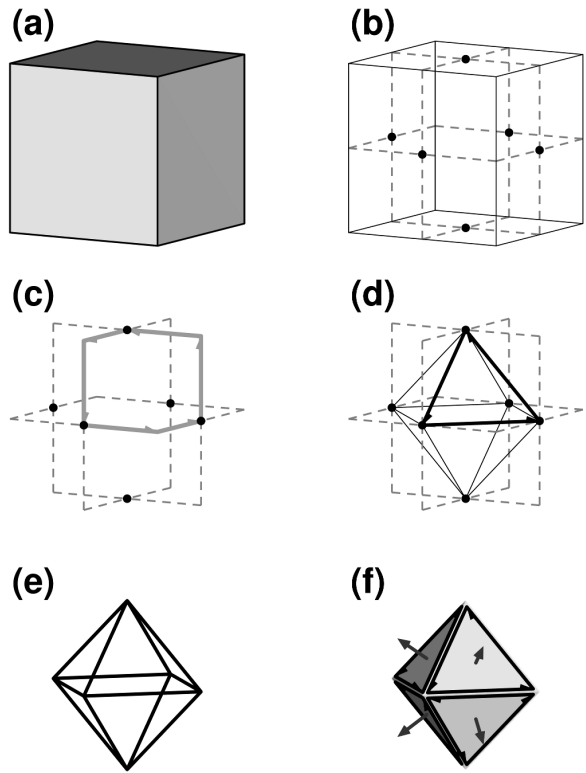}
\vspace{-0.1in} 
\caption{
(a) A single voxel (gray cube); 
(b) the six boundary faces of the single voxel (\textit{cf.}, Fig.~2.b);
(c) six voxel face vectors form a single vector cycle; 
(d) replacement of voxel face vector pairs with VESTA cycle vectors
(\textit{cf.}, Fig.s~3.f and~3.g);
(e) resulting wire frame after the processing of all voxel face vectors; 
(f) an octahedron, the corresponding final VESTA surface, which is 
superimposed with oriented $3$-cycles and normal vectors for the
visible gray triangles.
} \label{multib_04}
\end{figure}
\begin{figure*}
\epsfig{width=12.0cm,figure=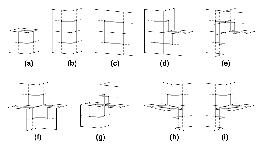}
\vspace{-0.1in} 
\caption{
Nine boundary face chains with their corresponding voxel face vectors;
note, that the directions of the vectors have been omitted, because
it is not specified on which side of the boundary faces the active
voxels reside;
in (e) and (g), the white dots mark junctures that are points of ambiguity.
} \label{multib_05}
\end{figure*}

Two pixels - one with a gray-level value of $133$, the other one with a zero 
valued gray-level - are initially separated by a contour section that is 
located exactly in the middle between them (dotted line). 
A range vector connects the centers of the two pixels. 
The centers of each pixel are assumed to correspond exactly 
with their gray-level values. 
Since the isocontour is supposed here to represent a gray-level of value $100$, 
it should not be positioned at the middle of the range vector. 
This medium position actually represents a gray-level of value $66.5$ 
while assuming a linear interpolation between the gray-level bounds. 
In fact, the ``true'' location of the support point for the isocontour is
located closer - and therefore has to be shifted - towards the center of the 
pixel with the gray-level of value $133$. 
Hence, the isocontour (solid line) is supported by a point, which is located 
within the pixel with the gray-level of value $133$.\\
\indent
In Fig.~2.d, we show the dislocation of the boundary face center in 3D due to
the previous outline for 2D\footnote{More complicated types of interpolations 
than just simple linear interpolations are possible.
E.g., one could use (higher-dimensional) B-Splines\cite{BOEH07, SALO05} 
instead, etc.}.
Note that the dashed gray lines each continue to provide a connection from the
boundary face center to one of the four edge middle points of the boundary face. 
In the following, these four edge middle points, which are numbered
counterclockwise from $1$ to $4$ in Fig.~2.b (i.e., if we apply a
right-hand-rule to the perpendicular range vector), will assume the role of 
junctures (see below). 
The following ansatz will help to provide a connectivity among the voxel face
vectors (VFV) within a given boundary face. 
Let us unite the eight black VFVs of Fig.~2.b into four vector pairs as 
follows:
connect the juncture $4$ ($3, 2, 1$) via the face center to the juncture $3$
($2, 1, 4$).
As a consequence, one obtains for each boundary voxel face the four
internal paths $4\rightarrow3$, $3\rightarrow2$, $2\rightarrow1$, and
$1\rightarrow4$, respectively.\\
\indent
Any single isolated active voxel or any active voxel cluster will initially
be fully enclosed by a certain number of boundary faces.
Hence, each boundary face will be in contact with at least another boundary
face through one of its four edges or -- to be more specific -- through one
of its four junctures.
Without loss of generality, we shall discuss in the following how pairs of
boundary faces may connect in 3D.
In Fig.~3, the various possible configurations are shown.
Through each juncture (white dots) two VFV pairs may be connected
(\textit{cf.}, Fig.s~3.a, 3.f, and~3.j).
While doing so, it is only permitted to attach the origin of a given VFV to
another ones tip. 
Hence, these vector pairs will yield oriented paths from one boundary 
face center to another one.\\
\indent
If we ignore the junctures, the newly formed vector pairs can be replaced
by single vectors, which each will connect one VESTA surface support point
with another one (\textit{cf.}, Fig.s~3.b, 3.g, and~3.k, respectively).
In the following, the latter vectors will be called ``VESTA cycle vectors''.
Throughout this paper, vector pairs of antiparallel vectors will
sometimes simply be drawn as single lines (\textit{cf.}, Fig.s~3.c, 3.d, 3.e,
3.h, 3.i, 3.l, and~3.m).
Furthermore, it may not be specified in one of the following drawings, on 
which side of a given boundary face the active voxel resides.
Then again, single vectors may be drawn as well as single line segments
where no particular orientation will be indicated, since it should be obvious
from the particular context.

\subsection{VESTA - a {\underline V}olume-{\underline E}nclosing 
{\underline S}urface ex{\underline T}raction {\underline A}lgorithm}

\noindent
In the previous subsection, we have discussed the basic 
processing steps that will allow for a proper surface extraction.
We shall begin with the processing of a single active, isolated voxel 
(\textit{cf.}, Fig.~4.a).
In Fig.~4.b, this voxel is represented by its six boundary faces.
As an example, only for three of the boundary faces, and only for a single
quadrant of each of these three faces, those VFVs are shown in Fig.~4.c, which
form a closed vector cycle. 
On the one hand, Fig.~2.b has made clear that VFVs are pairwise connected by 
the boundary face centers; 
on the other hand, Fig.~3.f suggests, that VFVs can be pairwise connected
through the junctures.\\
\indent
After the replacement of VFV pairs that are connected through junctures by the
proper VESTA cycle vectors (\textit{cf.}, Fig.~3.g), one obtains in Fig.~4.d a 
closed VESTA $3$-cycle, which represents a single oriented triangle.
The processing of all VFVs yields seven further triangles.
As a result, one obtains a fully closed and oriented VESTA surface
(\textit{cf.}, the octahedrons in Fig.s~4.e and~4.f).
In Fig.~4.f, the octahedron is superimposed with the four normal vectors of 
the four visible surface triangles.
These normal vectors point to the exterior of the enclosed shape.\\
\indent
All initially given $48$ VFVs are unique.
In total, one obtains after their processing $24$ VESTA cycle vectors
for the $12$ edges of the octahedron, i.e., each edge represents two 
antiparallel vectors.
Note, that VESTA will always reproduce 2D DICONEX contours whenever a final 
3D VESTA surface is intersected with a principal plane at the corresponding 
centers of the active voxels (\textit{cf.}, Fig.s~31.a --~31.c).

\begin{figure}[!hb]
\epsfig{width=8.0cm,figure=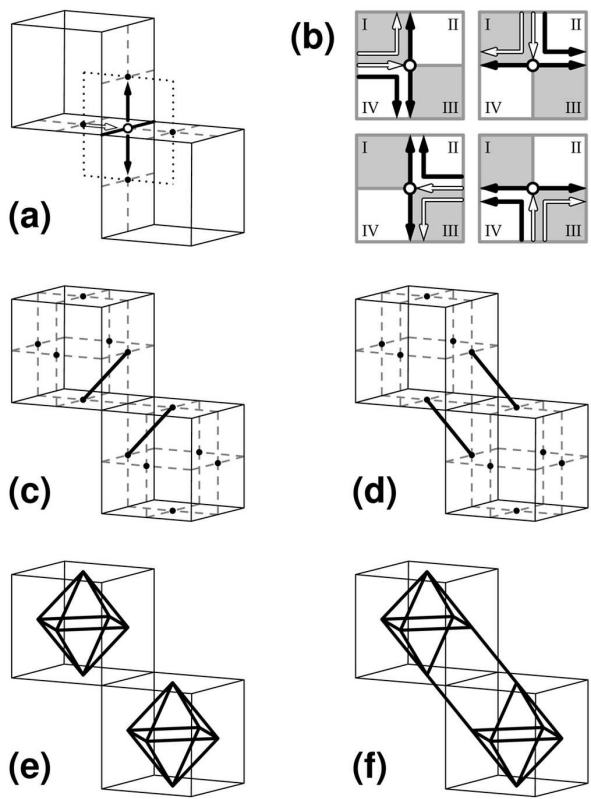}
\vspace{-0.1in} 
\caption{
Two voxels in contact through one single voxel edge
(a) the dotted square contains a single point of ambiguity (white dot)
with one incoming (white) and two outgoing (black) voxel face vectors;
(b) four connection diagrams for a single point of ambiguity (white dot);
(c) with VESTA cycle edges for the ``disconnect mode'', 
(d) with VESTA cycle edges for the ``connect'' mode; final wire frames for 
(e) the surfaces in ``disconnect'' mode, and 
(f) the surface in ``connect'' mode.
} \label{multib_06}
\end{figure}

In general, one has to process more complex shapes than just single voxels.
Let us assume for the remainder of this subsection that the 3D data set
is simply binarized (as in ``active'' and ``inactive'' voxels).
The more general situation will be discussed in one of the next subsections.
As in 2D, we shall encounter 3D junctures in the process of VESTA surface
cycle formation that will play the role of POAs.
In Fig.~5, we show all possible nine configurations of boundary faces that
can be in direct contact.
Gray solid lines represent single VFVs in the figure.
Most of the configurations show paths, which do not self-intersect.
However, in Fig.s~5.e and~5.g, we observe self-intersections, which are
due to the fact that two voxels are in contact with one another through only 
one single edge (\textit{cf.}, Fig.~6.a).\\
\begin{figure*}[t]
\epsfig{width=12.0cm,figure=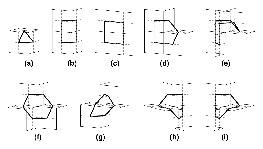}
\vspace{-0.1in} 
\caption{
Nine VESTA surface cycles with corresponding boundary face chains: 
(a) (planar) 3-cycle, 
(b) \& (c) planar 4-cycles, 
(d) nonplanar 5-cycle, 
(e) nonplanar 7-cycle, 
(f) planar 6-cycle, 
(g) -- (i) nonplanar 6-cycles.
} \label{multib_07}
\end{figure*}
\begin{figure}[b]
\epsfig{width=8.5cm,figure=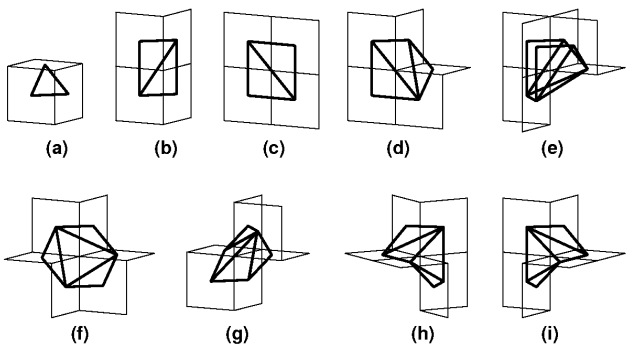}
\vspace{-0.1in} 
\caption{
Nine VESTA surface cycles with decomposition into triangles: 
(a) 3-cycle, 
(b) \& (c) 4-cycles, 
(d) 5-cycle, 
(e) 7-cycle, 
(f) -- (i) 6-cycles. 
No additional surface support points are used.
} \label{multib_08}
\end{figure}
\begin{figure}[b]
\epsfig{width=8.5cm,figure=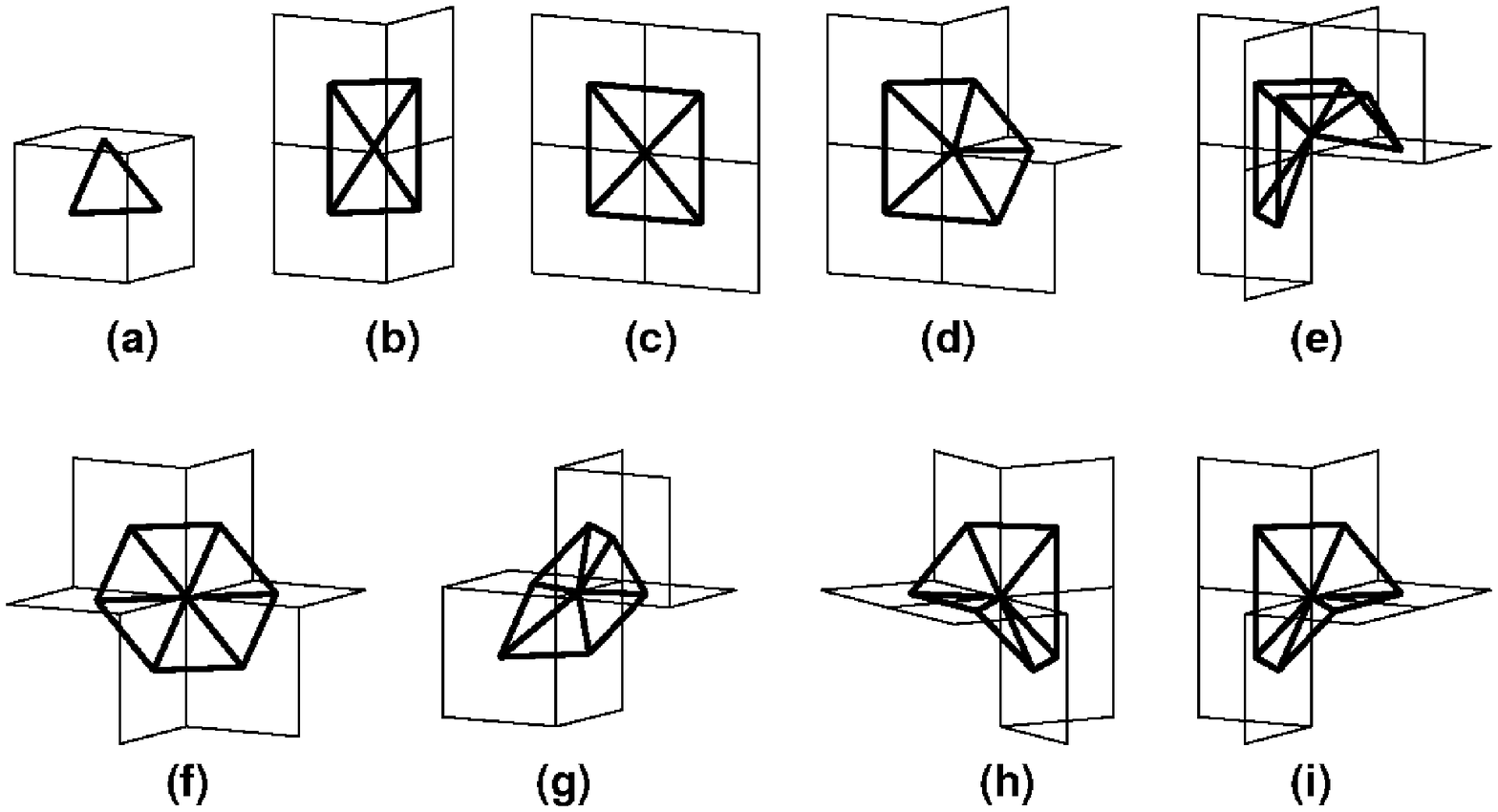}
\vspace{-0.1in} 
\caption{
Nine VESTA surface cycles with decomposition into triangles: 
(a) 3-cycle, 
(b) \& (c) 4-cycles, 
(d) 5-cycle, 
(e) 7-cycle, 
(f) -- (i) 6-cycles. 
For cycles with $N > 3$, average points are used to generate 
a surface tiling with triangles.
} \label{multib_09}
\end{figure}
Fig.~6.a also shows a dotted square, which has the common juncture of the 
four boundary faces that are in direct contact at its center. 
As an example, one incoming and two outgoing VFVs are shown for this juncture.
The upper left connection diagram of Fig.~6.b shows this configuration once
again.
However, due to the 3D nature of the problem we actually have four incoming
and four outgoing VFVs at this particular juncture.
Clearly, this juncture is a POA, and Fig.~6.b shows for each of the four
incoming VFVs a connection diagram with two valid outgoing VFVs.
In order to avoid holes in the final set of surface tiles, one has now
uniformly to choose among the two following possibilities for all four
connection diagrams.\\
\indent
Either, one selects as successor to a given VFV the one that is connected 
with its origin and that belongs to the same voxel; 
then one generates the ``disconnect'' mode while following the paths of the
white bent vectors (\textit{cf.}, Fig.~6.c). 
Or, one selects as a successor the one that is connected with its origin 
and that belongs to the other voxel; 
then one generates the ``connect'' mode while following the paths of the
black bent vectors (\textit{cf.}, Fig.~6.d). 
This selection process is the 3D analog of the 2D selection process 
(\textit{cf.}, Fig.~1.d).
A consistent selection of these successors is the key step within VESTA, which 
prevents \textit{a tearing of holes} into the final surface.
The user has to perform a global selection of one of the two modes, if only binary 
data are available.
In Fig.s~6.e and~6.f, the final VESTA surface wire frames are depicted for
the ``disconnect'' mode and for the ``connect'' mode, respectively.
Note, that each edge of the wire frames consists of two antiparallel
VESTA surface cycle vectors.\\
\indent
In Fig.~7, all possible nine VESTA surface cycles are shown (\textit{cf.},
Fig.~5), which can be generated while uniformly either choosing the 
``disconnect'' or the ``connect'' mode for the whole 3D data set under
consideration. 
Note, that these planar and nonplanar cycles can be traversed both ways,
depending on the orientation of the range vectors (not shown here) of the
contributing active voxels.
In particular, the cycles in Fig.s~7.e and~7.g are the VESTA surface cycles
of maximum length;
their separated counterparts can be seen in Fig.s~13.b and~13.a,
respectively. 
Furthermore, and as an important result, all of the cycles are confined to 
a $2\times2\times2$ voxel neighborhood at all times.
They are supported alone by the initial boundary face centers, which may vary
-- if necessary -- within the bounds defined by their corresponding range
vectors (\textit{cf.}, Fig.~2).
Finally, it should be noted that more than one VESTA surface cycle can
appear within a $2\times2\times2$ voxel neighborhood 
(\textit{cf.}, Fig.~13).

\subsection{VESTA Surface Cycle Decomposition}

\noindent
The VESTA surface cycles, which have been introduced in the previous
subsection, have to be processed further, if one wants to obtain surfaces,
which are decomposed into triangles.
In fact, the only cycle that requires no further processing is the
one that is shown in Fig.~7.a.
If we demand to use no further points while inserting edges, we may end
up with a result that is shown in Fig.~8.
Note that instead of inserting single edges, one rather has to insert 
antiparallel vector pairs for the proper breakup of the $N$-cycles ($N > 3$) 
into $3$-cycles.
This allows one to pass on the initial orientation of the $N$-cycles to the 
newly formed $3$-cycles.\\
\indent
In Fig.~8, the triangular surface tiles have been chosen similar to the 
template tiling, which has been proposed by various authors of the
MCA (\textit{cf.}, Ref.s~\cite{LORE87} and~\cite{BOUR94}). 
However, this tiling is not unique, since other subdivisions are possible.
If no further points are used for the decomposition of VESTA $N$-surface
with $N > 3$, biases may be introduced with respect to the convexity 
and/or concavity of the local surface sections (\textit{cf.}, e.g.,
Fig.s~18.e and~18.f). 
In cases, when more numerical accuracy is desired, it may be advisable to
introduce for each $N$-cycle with $N > 3$ an additional point that lies within 
the cycle (\textit{cf.}, e.g., Fig.s~18.c and~18.d).\\
\indent
In Fig.~9, the $N$-cycles ($N > 3$) of Fig.~7 have been broken down into
$3$-cycles (i.e., oriented triangles) while using as additional point the
average of all involved cycle support points.
Note that in the case where the initial boundary face centers are moved
within the bounds of the range vectors, the average point should be determined 
after this movement in order to save computing time.
Admittedly, this choice of the average is a simple one, but it will be used
for the remainder of this paper.
Ultimately, it is left to the designer of a particular VESTA implementation, 
whether a more elaborated approach should be used for the breakup of the 
surface cycles.

\subsection{More on Voxel Connectivity}

\noindent
In the case, that two neighboring active voxels only share a single point
(\textit{cf.}, Fig.~10.a), VESTA will not connect the two volumes, since none 
of the involved VFVs of one voxel will ever meet one of the VFVs of the 
other voxel. 
In Fig.~10.b, we show the wire frames (i.e., two octahedrons) for
this configuration.
Apparently, this connection is in the discretized 3D space too weak that it
should matter. 
If one -- nonetheless -- desires to establish a link between voxels that only 
share a single common point one may, e.g., insert a tubular template 
(while removing the two opposite triangles, i.e., $3$-cycles, at the same time)
as indicated through the six dashed lines in the figure.
However, we shall not use such an approach here, since VESTA by itself does 
not establish this kind of connectivity.

\begin{figure}[t]
\epsfig{width=8.5cm,figure=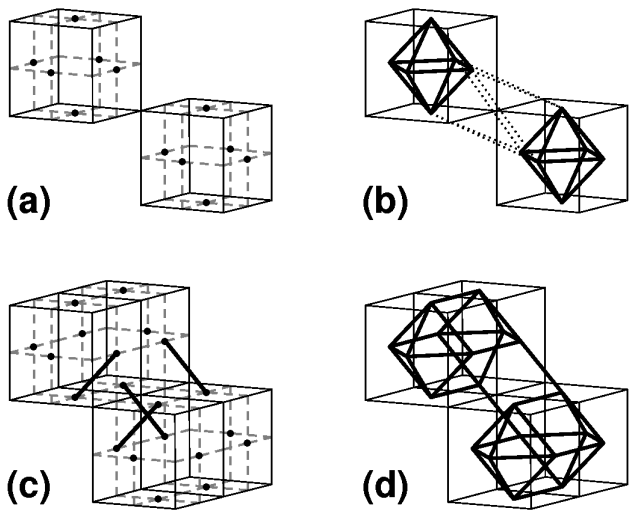}
\vspace{-0.1in} 
\caption{
(a) Two active voxels that have only a single point in common;
(b) the VESTA wire frame for the voxels shown in (a), together with
a connecting triangular tube template (dotted lines);
(c) two active voxel pairs in direct contact, which each just have a single 
edge in common, but with different connectivity modes;
(d) the wire frame for the voxels shown as in (c).
} \label{multib_10}
\end{figure}

So far, we have discussed only the uniform and global usage of either the
``disconnect'' or the ``connect'' mode for the processing of the 3D data 
set under consideration.
However, if more than just plain binary information (such as ``do enclose''
and ``do not enclose a voxel with a surface'') is contained in the data, e.g.,
gray-level information, we can {\it locally} define through a threshold, 
which mode should be (consistently) applied.
Ultimately, the interaction of the user is no longer required, since the 
thresholding could now be applied through automation.\\
\indent
In Fig.~10.c, we show two pairs of active voxels, which each just share a 
single edge, and that are in direct contact to each other.
Here, we consider the case that for each of the two neighboring pairs a 
different connectivity mode has been selected as it is indicated in the figure.
As a consequence of the tracing of VFVs (\textit{cf.}, Fig.s~11.a), a
new VESTA surface cycle comes into existence. 
This $8$-cycle is shown in Fig~12.a. 
The corresponding wire frame of the two voxel pairs is depicted in 
Fig~10.d where each edge of the wire frame consists of two antiparallel 
VESTA surface cycle vectors.
Note, that if both connection modes would have been the same, e.g.,
``connect mode'', we would have obtained the two VESTA $4$-cycles as
shown in Fig.~13.c.
In Fig.~14.a, the newly formed $8$-cycle is decomposed into $3$-cycles while
using an additional average point of the involved cycle support points.\\
\indent
In fact, many more VESTA surface cycles can appear, dependent on the internal
features of the considered 3D data.
All of these are shown in Fig.~12;
and in Fig.~14, we show their decomposition into triangles while using average 
points.
In contrast to VESTA, the original MCA~\cite{LORE87} \textit{does not} propose 
templates for the here described further configurations.
However, various extensions (\textit{cf.}, Ref.s~\cite{CHER95,LEWI03,NEWM06}) of
the MCA use additional templates, which also use an additional average point 
for their triangular decomposition.
In general, there is now a total of $14$ different types of VESTA $N$-cycles 
($N=3, 4, 5, 6, 7, 8, 9, 12$).
In particular, $N$-cycles with a length of $10$ or $11$ do not exist,
since (complementary) $N$-cycles with a length of $2$ or $1$, respectively, 
do not exist either.

\begin{figure}[t]
\epsfig{width=8.5cm,figure=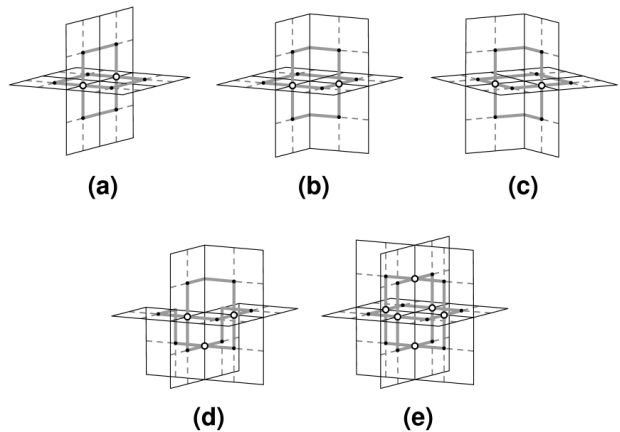}
\vspace{-0.1in} 
\caption{
Four further boundary face chains with their corresponding voxel face vectors;
the directions of the vectors have been omitted, because it is not specified on
which side of the boundary faces the active voxels reside;
the white dots represent points of ambiguity;
note, that figures (b) and (c) are equal, but they will eventually lead
to different VESTA surface cycles, and therefore, they are drawn twice here.
} \label{multib_11}
\end{figure}
\begin{figure}[b]
\epsfig{width=8.5cm,figure=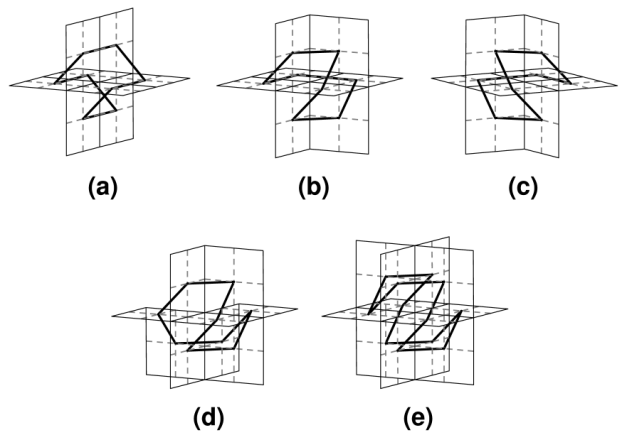}
\vspace{-0.1in} 
\caption{
Five more VESTA surface cycles (of maximum length) with corresponding
boundary face chains: 
(a) -- (c) nonplanar 8-cycles, 
(d) nonplanar 9-cycle, 
(e) nonplanar 12-cycle. 
} \label{multib_12}
\end{figure}

\begin{figure}[t]
\epsfig{width=9.0cm,figure=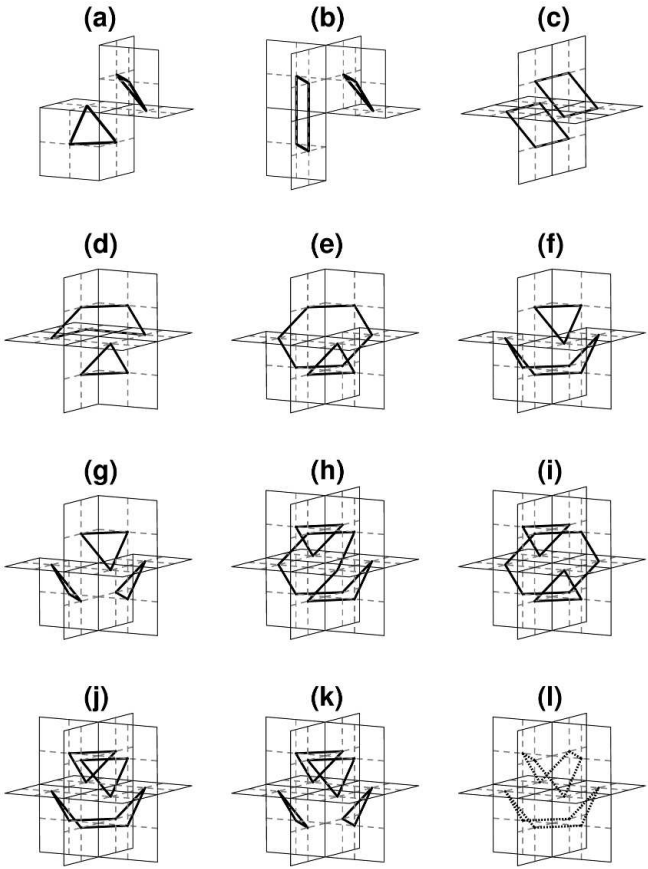}
\vspace{-0.1in} 
\caption{
Generation of multiple VESTA surface cycles:
(a) results from Fig.~5.g;
(b) results from Fig.~5.e;
(c) results from Fig.~11.a;
(d) results from Fig.~11.b or~11.c;
(e) -- (g) result from Fig.~11.d;
(h) -- (k) result from Fig.~11.e;
(l) can never be created (Ref.~\cite{CHER95} provides 
an exhaustive discussion).
} \label{multib_13}
\end{figure}
\begin{figure}[!b]
\epsfig{width=8.5cm,figure=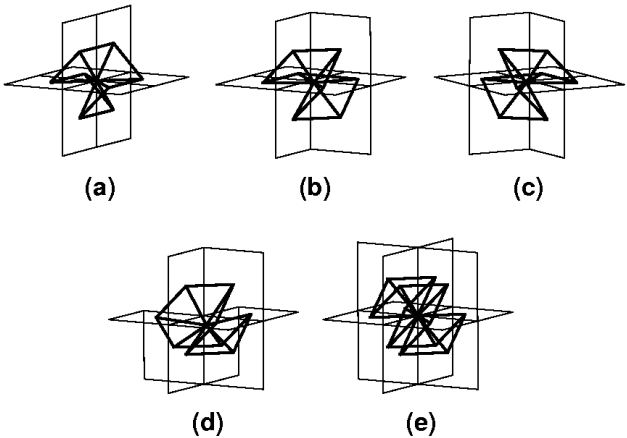}
\vspace{-0.1in} 
\caption{
The VESTA surface cycles as shown in Fig. 12 with decomposition into triangles: 
(a) -- (c) 8-cycles, 
(d) 9-cycle, 
(e) 12-cycle. 
Here, average points are used to generate a surface tiling with triangles.
} \label{multib_14}
\end{figure}

We would like to stress that VESTA can produce in total six different
types of surfaces.
There are the surfaces, which will be processed either in the global
``disconnect'' or ``connect'' modes.
Then only VESTA $N$-cycles of length up to seven will be generated.
A third kind is generated while using the ``mixed'' mode, which has been
described in this subsection.
Furthermore, each of the three modes can be either executed in a so-called
low resolution mode (``L'') where no additional points are used for the
decomposition of the surface cycles; 
or in a so-called high resolution mode (``H'') where additional average
points are used for the decomposition of surface cycles.
Up to this point, we have described what we will later refer to as the
``original VESTA''.
Note, that this type of VESTA has been implemented successfully into 
efficient software~\cite{VEST04}.
In the following subsection, we shall discuss a marching variant of VESTA.

\begin{figure}[!b]
\epsfig{width=8.5cm,figure=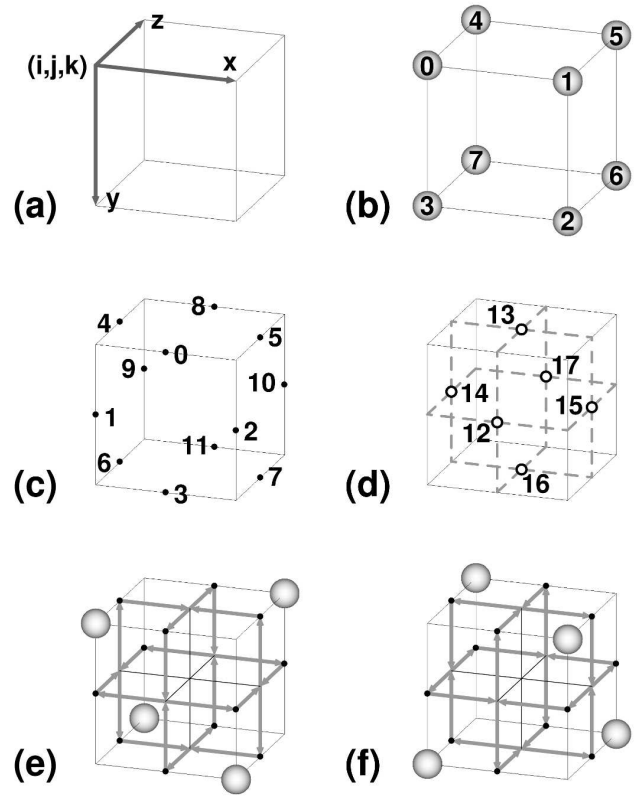}
\vspace{-0.1in} 
\caption{
Various conventions for the marching VESTA for a given $3$-cell:
(a) a marching $3$-cell at position $(i,j,k)$, superimposed with a
cartesian coordinate system, $x|y|z$;
(b) numbering of voxels; the spheres mark active voxel centers,
or ``sites'';
(c) numbering of VESTA surface support points, i.e., boundary
face centers;
(d) numbering of junctures, which may or may not become 
points of ambiguity;
the dashed lines represent pairs of anti-parallel voxel face vectors;
(e) voxel face vectors for voxels no.s $0$, $2$, $5$, and $7$;
(f) voxel face vectors for voxels no.s $1$, $3$, $4$, and $6$;
in (e) and (f) the black dots represent boundary face centers.
} \label{multib_15}
\end{figure}

\subsection{The Marching 3-Cell Version of VESTA}

\noindent
Using the original VESTA, one processes the 3D data set under consideration
at once.
This has the advantage that redundant information can be mostly avoided
during the generation of surface tiles.
However, a clear disadvantage is that a large amount of computer
memory may be required, in particular, if large amounts of voxels have to
be processed.
As a consequence, a computer with insufficient RAM may be forced to go into 
a swapping mode, and the execution time may increase considerably.
Therefore, we shall describe in the following an alternate implementation
of VESTA, which we will name the ``marching VESTA''.\\
\indent
In this subsection, we are going to describe the details of a scan of a given 
3D data set with a marching $2\times2\times2$ voxel neighborhood ($3$-cell, 
or cube).
All of the surface tiles, which finally will represent a single
or multiple VESTA surface(s), will be generated and collected during such a
scan. 
Note, that this scanning of the data is the same as in case of the 
MCA~\cite{LORE87,BOUR94}. 
However, we would like to stress, that we are confident to use such an 
approach only, because the ansatz of using the active voxels' boundary
faces as shown in Fig.~2.b has resulted in VESTA surface $N$-cycles, which 
are all confined to $3$-cells (\textit{cf.}, Fig.s~7 and~12).
In the following, we shall outline the necessary steps for a more memory
efficient VESTA implementation while using a few examples.\\
\begin{center}
\begin{table}[b]
\begin{tabular}{||r||c|c||}
\hline
\hline
\textbf{Center ID$\:$}
&{\boldmath$\oplus$}$\:$\textbf{Path}
&{\boldmath$\ominus$}$\:$\textbf{Path}\\
\hline
\hline
{\boldmath$0\:$}
&$\: 13 \rightarrow 0\rightarrow 12 \:$
&$\: 12 \rightarrow 0\rightarrow 13 \:$\\
\hline
{\boldmath$1\:$}
&$\: 12 \rightarrow 1\rightarrow 14 \:$
&$\: 14 \rightarrow 1\rightarrow 12 \:$\\
\hline
{\boldmath$2\:$}
&$\: 15 \rightarrow 2\rightarrow 12 \:$
&$\: 12 \rightarrow 2\rightarrow 15 \:$\\
\hline
{\boldmath$3\:$}
&$\: 12 \rightarrow 3\rightarrow 16 \:$
&$\: 16 \rightarrow 3\rightarrow 12 \:$\\
\hline
{\boldmath$4\:$}
&$\: 14 \rightarrow 4\rightarrow 13 \:$
&$\: 13 \rightarrow 4\rightarrow 14 \:$\\
\hline
{\boldmath$5\:$}
&$\: 13 \rightarrow 5\rightarrow 15 \:$
&$\: 15 \rightarrow 5\rightarrow 13 \:$\\
\hline
{\boldmath$6\:$}
&$\: 16 \rightarrow 6\rightarrow 14 \:$
&$\: 14 \rightarrow 6\rightarrow 16 \:$\\
\hline
{\boldmath$7\:$}
&$\: 15 \rightarrow 7\rightarrow 16 \:$
&$\: 16 \rightarrow 7\rightarrow 15 \:$\\
\hline
{\boldmath$8\:$}
&$\: 17 \rightarrow 8\rightarrow 13 \:$
&$\: 13 \rightarrow 8\rightarrow 17 \:$\\
\hline
{\boldmath$9\:$}
&$\: 14 \rightarrow 9\rightarrow 17 \:$
&$\: 17 \rightarrow 9\rightarrow 14 \:$\\
\hline
{\boldmath$10\:$}
&$\: 17 \rightarrow 10\rightarrow 15 \:$
&$\: 15 \rightarrow 10\rightarrow 17 \:$\\
\hline
{\boldmath$11\:$}
&$\: 16 \rightarrow 11\rightarrow 17 \:$
&$\: 17 \rightarrow 11\rightarrow 16 \:$\\
\hline
\hline
\end{tabular}
\caption{
Directed paths for the quadrants of the oriented boundary faces, which have
their centers at the predefined locations as depicted in Fig.~15.c.
The start and end points of the $24$ paths are the junctures, which are
shown in Fig.~15.d.
}
\label{table_01}
\end{table}
\end{center}
\vspace*{-1.2cm}
\indent
\indent
In Fig.~15.a, a marching $3$-cell at position $(i,j,k)$ is shown.
The indices $i$, $j$, and $k$, refer to the position of voxel no. $0$
(\textit{cf.}, Fig.~15.b) in the $x$-, $y$-, and $z$-directions,
respectively.
Note that the coordinate system, $x|y|z$, coincides with the range
vectors (\textit{cf.}, Fig.~2) of voxel no. $0$ for positive $x$-, 
$y$-, and $z$-directions.
The VESTA surface support points (\textit{cf.}, Fig.~15.c) can move within 
the bounds of the range vectors, which coincide with the edges of the marching
$3$-cell. 
Since an active voxel may show up in a marching $3$-cell only with an
eighth of its volume (i.e., an octant), its initial corresponding boundary 
faces may show up only as quadrants (\textit{cf.}, Fig.~2).\\
\indent
In Table~1, the paths of VFV pairs are listed, which each correspond to
a quarter (i.e., a quadrant) of a potential boundary face of an active
voxel within the $3$-cell $(i,j,k)$.
These paths connect junctures (\textit{cf.}, Fig.~15.d) via the boundary
face centers (\textit{cf.}, Fig.~15.c) for the given boundary faces.
Since each boundary face is shared by two side by side voxels, two 
orientations, $\oplus$ and $\ominus$, exist for the VFV based paths
with respect to the positive and negative spatial directions.
In Fig.s 15.e and~15.f together, all $48$ possible VFVs are shown for 
the marching $3$-cell.\\
\indent
As an example, in Fig.~16.a, a $3$-cell is shown with the single active
site, no. $7$, together with an octant of the active voxel, representing 
its partial volume. 
The three black dots in the figure represent the corresponding boundary 
face centers, and the six dashed lines connect these with the corresponding
junctures.
I.e., for a single active site, only three boundary face quadrants have
to be considered at maximum.
Note, that each of the boundary face quadrants are oriented.
In Fig.s~16.b, 16.c, and~16.d, the corresponding (gray) VFV pairs are drawn,
together with their corresponding boundary face centers,
i.e., the paths $9\ominus$, $6\ominus$, and $11\oplus$ (\textit{cf.}, 
Table~1), respectively.
The three paths connect to the closed point sequence, 
$17 \rightarrow 9\rightarrow 14 \rightarrow 6\rightarrow 16 \rightarrow
11\rightarrow 17$.
After the removal of points with identities $>11$ (i.e., junctures), the 
sequence will transform into the VESTA $3$-cycle, 
$9\rightarrow 6\rightarrow 11\rightarrow 9$.
As a result, we obtain a single oriented triangle within the $3$-cell.\\
\begin{figure}[t]
\epsfig{width=8.5cm,figure=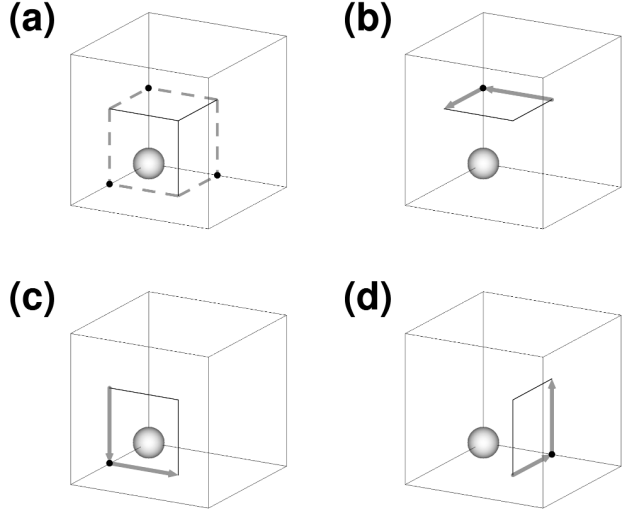}
\vspace{-0.1in} 
\caption{
A $3$-cell with one single active voxel (sphere, 
representing its voxel center) together with the
(a) corresponding octant of the active voxel volume,
(b) - (d) the three different, contributing boundary face quadrants, 
i.e., the boundary face centers (black dots) and two voxel face vectors
(gray) each.
} \label{multib_16}
\end{figure}
\begin{figure}[b]
\epsfig{width=8.5cm,figure=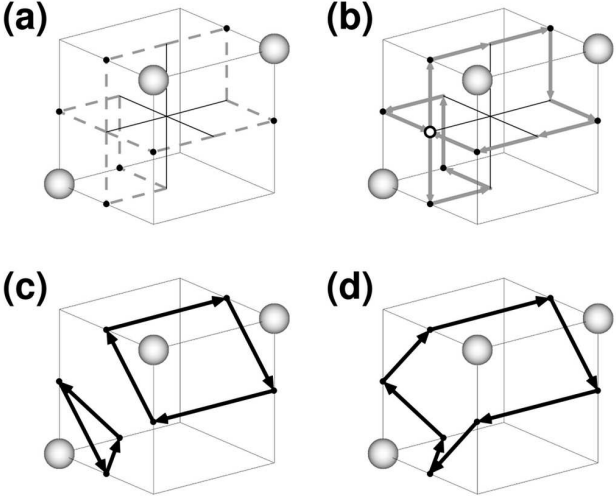}
\vspace{-0.1in} 
\caption{
A $3$-cell with three active voxels (spheres,
representing their voxel centers) together with
(a) the corresponding octants of the active voxel volumes,
(b) the various contributing boundary face quadrants, seven boundary face
centers (black dots), fourteen voxel face vectors (gray), and a single
point of ambiguity (white dot),
(c) a single $3$-cycle and a single $4$-cycle that are resulting from the
``disconnect'' mode, and
(d) a single $7$-cycle that results from the ``connect'' mode.
} \label{multib_17}
\end{figure}
\indent
\begin{figure}[t]
\epsfig{width=8.5cm,figure=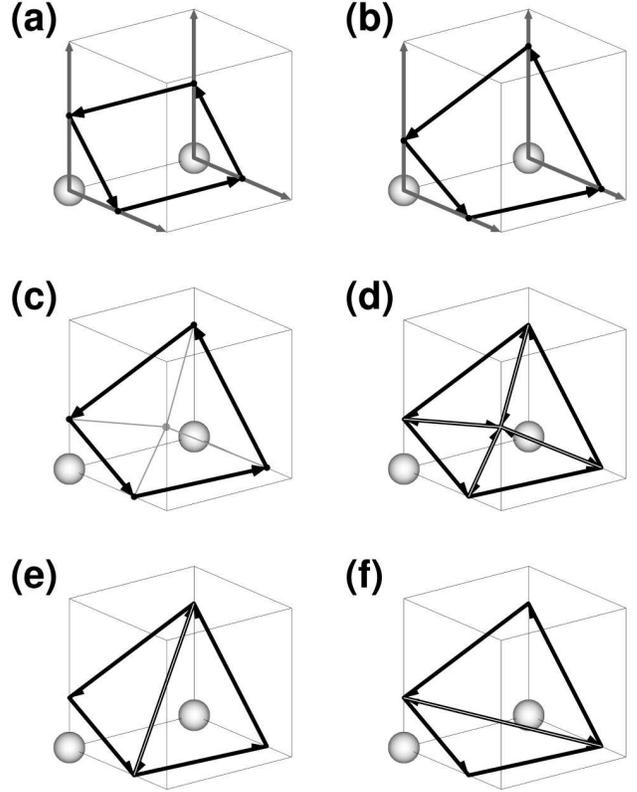}
\vspace{-0.1in} 
\caption{
(a) A $3$-cell with two active voxels (spheres, representing their voxel
centers) together with the corresponding range vectors and a resulting 
$4$-cycle;
(b) as in (a), but with displaced boundary face centers;
(c) as in (b), but without the range vectors and an additional average point
(gray) and further line segments (gray) indicating the anticipated triangle
decomposition;
(d) as in (c) with four $3$-cycles instead of a single $4$-cycle;
(e) a convex-shaped, and 
(f) a concave-shaped solution without the usage of additional support points. 
Note, that the orientations of the newly formed $3$-cycles are inherited from
the initially given $4$-cycle.
} \label{multib_18}
\end{figure}

\indent
In another example, we shall process the three active sites as shown in
Fig.~17. 
First, all of the contributing boundary face quadrants are determined
with their proper orientation.
In Fig.~17.a, the according seven boundary face quadrants are shown,
together with the corresponding boundary face centers.
In a next step, one considers all relevant VFV paths (\textit{cf.}, 
Fig.~17.b).
If we focus on the six faces of the $3$-cell, we notice that a face of a
$3$-cell can always contain either zero, two, or four VFVs.
In this given example, five of the six $3$-cell faces contain only
two VFVs each.
With the help of Table~1, the connection of the vectors is straightforward
here.\\
\indent
But, one $3$-cell face contains four VFVs, and at its center we observe
a POA.
Note that for the POA the application of the connection diagram as shown 
in Fig.~1.d will suffice for the proper execution of a previously specified
``disconnect'' and/or ``connect'' mode.
In other words, for the marching VESTA, the much more complicated usage
of the four connection diagrams as shown in Fig.~6.b is not necessary,
hence, lesser computational decisions have to be made.
Finally, considering the chosen connectivities among the VFVs, and after
omission of the junctures, we end up with two solutions for the VESTA surface
cycles.
We obtain either a single $3$-cycle and a single $4$-cycle that are resulting 
from the ``disconnect'' mode (\textit{cf.}, Fig.~17.c), or we obtain a 
single $7$-cycle that results from the ``connect'' mode (\textit{cf.},
Fig.~17.d).\\
\indent
For the sake of completeness, we demonstrate in Fig.~18 for two active sites 
the recommended processing steps for boundary face center displacement 
(which is in general necessary for isosurface generation, \textit{cf.},
Fig.s~18.a and~18.b), and the breakup of the VESTA $N$-cycles ($N > 3$) into
$3$-cycles (\textit{cf.}, Fig.s~18.c --~18.f). 
In this given example, we have one single VESTA $4$-cycle. 
Note, that the contributing range vectors coincide with the respective edges 
of the $3$-cell (\textit{cf.}, Fig.s~18.a and~18.b).
Furthermore, it is recommended that the boundary face centers will be 
displaced before the breakup into $3$-cycles, because then one is not 
required to reevaluate the 3D position of the possibly inserted average point, 
which once again may save computing time.
Fig.s~18.e and~18.f illustrate the introduction of biases for the shape
of the generated surface section, if no additional support points, e.g.,
average points, will be used.\\
\indent
This concludes the description of the marching VESTA.

\subsection{VESTA vs. the Marching Cubes Algorithm}

\begin{figure*}[pt]
\epsfig{width=13.0cm,figure=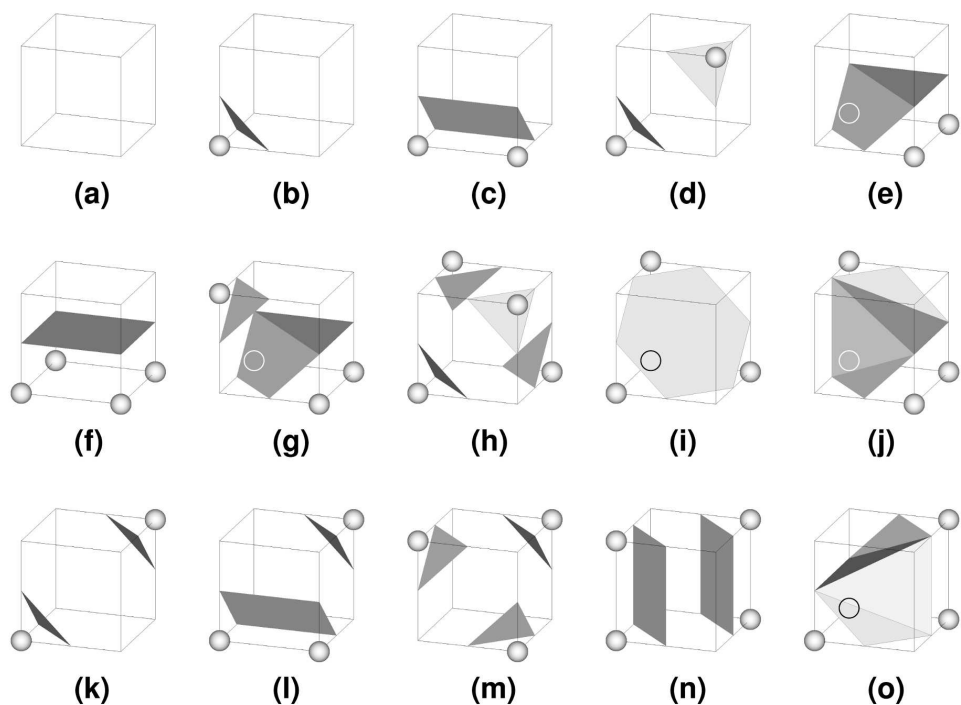}
\vspace{0.1in} 
\caption{
Full surface template set of the original Marching Cubes algorithm.
White and black circles indicate blanketed active $3$-cell sites.
} \label{multib_19}
\end{figure*}
\begin{figure*}[!pt]
\epsfig{width=13.0cm,figure=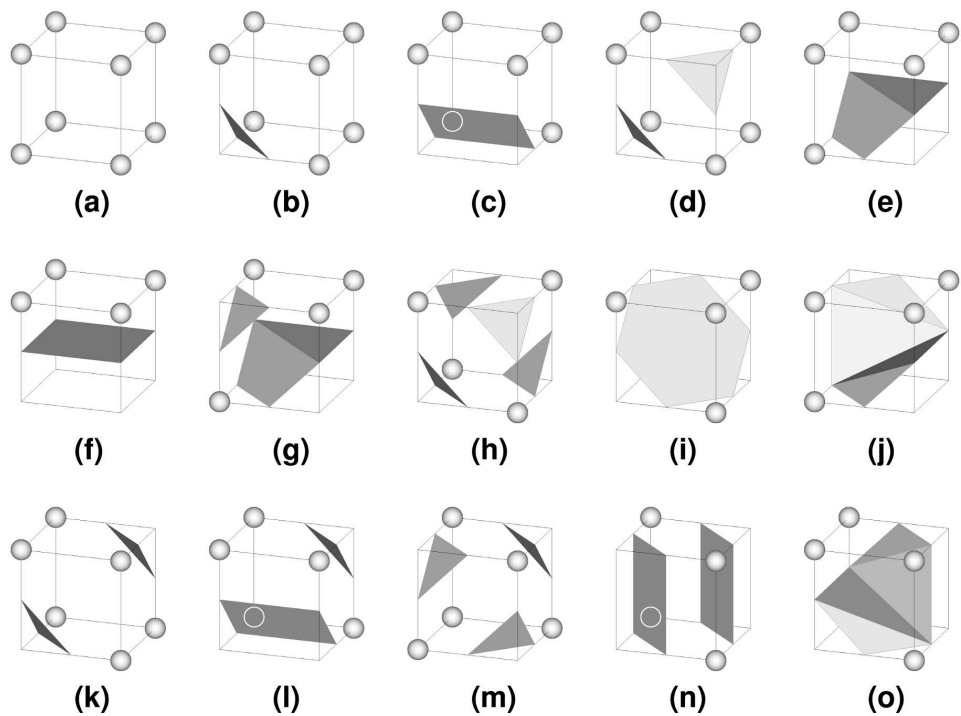}
\vspace{0.1in} 
\caption{
As in Fig.~19, but with inverted $3$-cell sites.
} \label{multib_20}
\end{figure*}
\begin{figure*}[pt]
\epsfig{width=13.0cm,figure=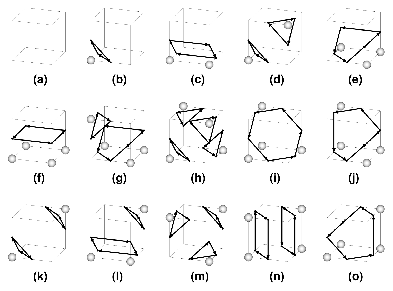}
\vspace{0.1in} 
\caption{
VESTA surface cycles in ``disconnect" mode,
for the $3$-cells shown in Fig.~19.
} \label{multib_21}
\end{figure*}
\begin{figure*}[!pt]
\epsfig{width=13.0cm,figure=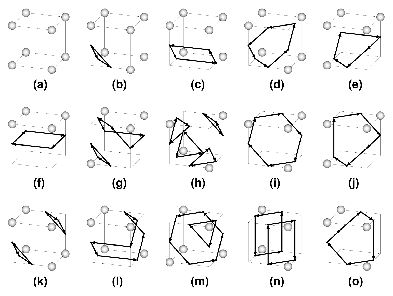}
\vspace{0.1in} 
\caption{
VESTA surface cycles in ``disconnect" mode,
for the $3$-cells shown in Fig.~20.
} \label{multib_22}
\end{figure*}
\begin{figure*}[pt]
\epsfig{width=13.0cm,figure=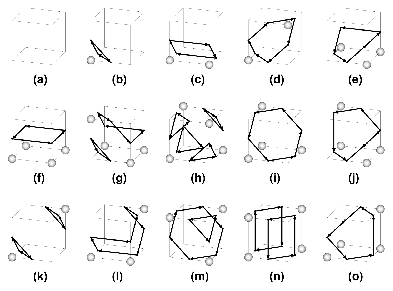}
\vspace{0.1in} 
\caption{
VESTA surface cycles in ``connect" mode,
for the $3$-cells shown in Fig.~19.
} \label{multib_23}
\end{figure*}
\begin{figure*}[!pt]
\epsfig{width=13.0cm,figure=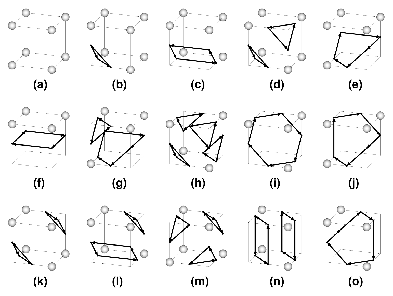}
\vspace{0.1in} 
\caption{
VESTA surface cycles in ``connect" mode,
for the $3$-cells as shown in Fig.~20.
} \label{multib_24}
\end{figure*}
\begin{figure}[t]
\epsfig{width=8.5cm,figure=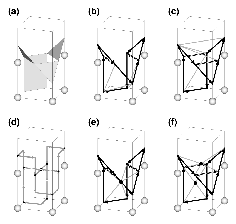}
\vspace{-0.1in} 
\caption{
Comparison between the Marching Cubes algorithm and VESTA for two
side-by-side $3$-cells:
(a) a hole is created due to the original MCA's limited template set;
(b) ``disconnect" mode solution for VESTA with low (``L'') resolution 
cycle decomposition;
(c) ``connect" mode solution for VESTA with L-decomposition;
(d) voxel face vectors (gray), boundary face centers (black dots) 
and a single point of ambiguity (white dot);
(e) as in (b), but with high (``H'') resolution cycle decomposition;
(f) as in (c), but with H-decomposition.
} \label{multib_25}
\end{figure}
\noindent
In this subsection, we provide an in-depth comparison between VESTA
and the original MCA\cite{LORE87}, because the original MCA can produce 
holes in the final surfaces, whereas VESTA does not.
Since the marching VESTA scans the given 3D data set exactly like the
MCA, we can compare the results of VESTA $3$-cell by $3$-cell with
the various templates, which the original MCA uses.
Since VESTA can distinguish between a ``disconnect'' and a ``connect''
mode whenever POAs are encountered, we shall present both solutions
for all given $3$-cells.\\
\indent
In Fig.~19, the complete surface section template set of the original 
MCA is shown for its $15$ configurations of different $3$-cell site 
occupancies (\textit{cf.}, also, e.g., Ref.s~\cite{LOHM98,NVID07}).
The original MCA uses the same set of templates for its correspondingly 
inverted $3$-cell sites (\textit{cf.}, Fig.~20).
In comparison to Fig.~19, we show the computed surface $N$-cycles for
VESTA in global ``disconnect'' (``connect'') mode in Fig.~21 (Fig.~23);
and in comparison to Fig.~20, we show the computed surface $N$-cycles for
VESTA in global ``disconnect'' (``connect'') mode in Fig.~22 (Fig.~24).
Several differences can be observed.\\
\indent
First, we provide a comparison between VESTA in its global ``disconnect'' 
mode and the original MCA.
VESTA reproduces in Fig.~21 perfectly the perimeters of the MCA template
set (\textit{cf.}, Fig.~19).
However, if we compare Fig.~22 with Fig.~20, there are differences
between the cases (d), (g), (h), (l), (m), and (n).
In particular, the original MCA does not provide templates like the
processed VESTA $N$-cycles shown in Fig.s~8.e and~8.g.\\
\indent
Secondly, we provide a comparison between VESTA in its global ``connect'' 
mode and the original MCA.
VESTA reproduces in Fig.~24 perfectly the perimeters of the MCA template
set for inverted sites (\textit{cf.}, Fig.~20).
However, if we compare Fig.~23 with Fig.~19, there are differences
-- once again -- between the cases (d), (g), (h), (l), (m), and (n).
As already pointed out, the original MCA misses templates 
like the processed VESTA $N$-cycles shown in Fig.s~8.e and~8.g.\\
\indent
Finally, we show in Fig.~25.a an example where the combination of the
original MCA's templates leads to the creation of a hole in the surface.
On the contrary, VESTA consistently applies the ``disconnect'' and/or
``connect'' modes for the two side-by-side $3$-cells under consideration,
and it does not create any holes (\textit{cf.}, Fig.s~25.b and~25.c).
The reason for VESTA's success lies in the ansatz based on the VFVs
of Fig.~2.b, and in the careful resolution of ambiguities at the junctures
(\textit{cf.}, Fig.~25.d).
In Fig.s~25.e and~25.f, VESTA's high resolution surface tilings are shown
in addition for this particular example.\\
\indent
Apparently, it is wrong to just take the same template set also for 
inverted $3$-cells.
Nowadays, various extensions of the original MCA account for the
missing surface templates (\textit{cf.}, e.g., 
Ref.s~\cite{CHER95,NEWM06,BOUR94}). 
In fact, some authors seem to provide too many additional surface 
templates (\textit{cf.}, Ref.~\cite{LEWI03}), i.e., some of these are 
actually not necessary for an adequate surface extraction.\\
\indent
Instead of using many surface templates, VESTA only uses a single
building block (\textit{cf.}, Fig.~2.b) as the basis for surface 
construction.
This concludes the theoretical section of this paper.\\

\begin{figure*}[pt]
\epsfig{width=12.0cm,figure=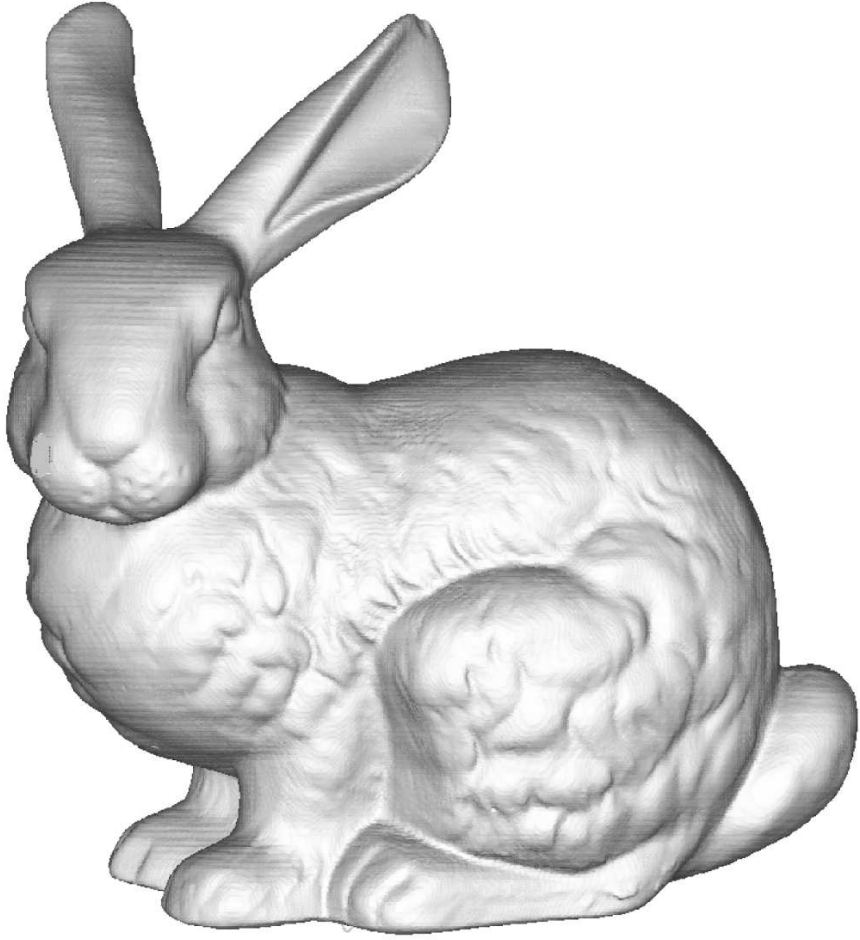}
\vspace{-0.1in} 
\caption{
VESTA isosurface rendering for CT-scan data of the Stanford Terra-Cotta
Bunny~\cite{STAN01}.
} \label{multib_26}
\end{figure*}
\begin{center}
\begin{table*}[pht]
\begin{tabular}{||r||c|c|c||c|c|c||c||}
\hline
\hline
\textbf{Technique}
&\multicolumn{3}{|c||}{original VESTA}
&\multicolumn{3}{|c||}{marching VESTA}
&$\:$extended MCA$\:$\\
\hline
\hline
\textbf{Connectivity$\:$}
&\textbf{Disconnect}&\textbf{Connect}&\textbf{Mixed}
&\textbf{Disconnect}&\textbf{Connect}&\textbf{Mixed}
&\textbf{Disconnect}\\
\hline
\hline
{\boldmath$3$}\textbf{-Cycles}$\:$
&$\: 254,662 \:$
&$\: 254,662 \:$
&$\: 254,662 \:$
&$\: 254,662 \:$
&$\: 254,662 \:$
&$\: 254,662 \:$
&$\: 254,662 \:$\\
\hline
{\boldmath$4$}\textbf{-Cycles}$\:$
&$\: 550,229 \:$
&$\: 550,229 \:$
&$\: 550,229 \:$
&$\: 550,229 \:$
&$\: 550,229 \:$
&$\: 550,229 \:$
&$\: 550,229 \:$\\
\hline
{\boldmath$5$}\textbf{-Cycles}$\:$
&$\: 178,512 \:$
&$\: 178,512 \:$
&$\: 178,512 \:$
&$\: 178,512 \:$
&$\: 178,512 \:$
&$\: 178,512 \:$
&$\: 178,512 \:$\\
\hline
{\boldmath$6$}\textbf{-Cycles}$\:$
&$\: 38,063 \:$
&$\: 38,063 \:$
&$\: 38,063 \:$
&$\: 38,063 \:$
&$\: 38,063 \:$
&$\: 38,063 \:$
&$\: 38,063 \:$\\
\hline
{\boldmath$7$}\textbf{-Cycles}$\:$
&$\: 0 \:$
&$\: 0 \:$
&$\: 0 \:$
&$\: 0 \:$
&$\: 0 \:$
&$\: 0 \:$
&$\: 0 \:$\\
\hline
{\boldmath$8$}\textbf{-Cycles}$\:$
&$\: N/A \:$
&$\: N/A \:$
&$\: 0 \:$
&$\: N/A \:$
&$\: N/A \:$
&$\: 0 \:$
&$\: N/A \:$\\
\hline
{\boldmath$9$}\textbf{-Cycles}$\:$
&$\: N/A \:$
&$\: N/A \:$
&$\: 0 \:$
&$\: N/A \:$
&$\: N/A \:$
&$\: 0 \:$
&$\: N/A \:$\\
\hline
{\boldmath$12$}\textbf{-Cycles}$\:$
&$\: N/A \:$
&$\: N/A \:$
&$\: 0 \:$
&$\: N/A \:$
&$\: N/A \:$
&$\: 0 \:$
&$\: N/A \:$\\
\hline
\textbf{Cycle Sum}$\:$
&$\: 1,021,466 \:$
&$\: 1,021,466 \:$
&$\: 1,021,466 \:$
&$\: 1,021,466 \:$
&$\: 1,021,466 \:$
&$\: 1,021,466 \:$
&$\: 1,021,466 \:$\\
\hline
\hline
\textbf{L: Points}$\:$
&{\boldmath$\: 1,021,460 \:$}
&$\: 1,021,460 \:$
&$\: 1,021,460 \:$
&{\boldmath$\: 4,085,840 \:$}
&$\: 4,085,840 \:$
&$\: 4,085,840 \:$
&{\boldmath$\: 6,128,724 \:$}\\
\hline
\textbf{L: Triangles}$\:$
&$\: 2,042,908 \:$
&$\: 2,042,908 \:$
&$\: 2,042,908 \:$
&$\: 2,042,908 \:$
&$\: 2,042,908 \:$
&$\: 2,042,908 \:$
&$\: 2,042,908 \:$\\
\hline
\textbf{L: Time [s]}$\:$
&{\boldmath$\: 20.44(7) \:$}
&$\: 20.48(5) \:$
&$\: 20.38(2) \:$
&{\boldmath$\: 29.62(34) \:$}
&$\: 29.62(31) \:$
&$\: 29.51(30) \:$
&{\boldmath$\: 28.20(41) \:$}\\
\hline
\hline
\textbf{H: Points}$\:$
&$\: 1,788,264 \:$
&$\: 1,788,264 \:$
&$\: 1,788,264 \:$
&$\: 4,852,644 \:$
&$\: 4,852,644 \:$
&$\: 4,852,644 \:$
&$\: N/A \:$\\
\hline
\textbf{H: Triangles}$\:$
&$\: 3,576,516 \:$
&$\: 3,576,516 \:$
&$\: 3,576,516 \:$
&$\: 3,576,516 \:$
&$\: 3,576,516 \:$
&$\: 3,576,516 \:$
&$\: N/A \:$\\
\hline
\textbf{H: Time [s]}$\:$
&$\: 20.44(2) \:$
&$\: 20.88(2) \:$
&$\: 20.86(7) \:$
&$\: 30.86(44) \:$
&$\: 30.30(45) \:$
&$\: 30.44(46) \:$
&$\: N/A \:$\\
\hline
\hline
\end{tabular}
\caption{
Stanford Terra-Cotta Bunny CT-scan data benchmark: 
$361$ images with dimensions $512\times512$ = $262,144$ pixels each; 
isovalue equals to $150$; 
number of active voxels: $3,100,197$, i.e., $3.276\%$.
}
\label{table_02}
\end{table*}
\end{center}

\vspace*{-1.0cm}
\section{Applications}

\noindent
In this application section, VESTA is used to create isosurfaces for
3D image data that have been generated from CT-scans~\cite{HERM09} and 
x-ray microtomography~\cite{STOC08}, respectively.
Furthermore, it is used to create a freezeout hypersurface
from a set of 3D numerical simulation data~\cite{ORNI92, BRS93} that 
result from the field of theoretical relativistic heavy-ion physics.
We also provide benchmarks between the original VESTA, the marching
VESTA, and an extended computer code implementation~\cite{BOUR94} of 
the original MCA (\textit{cf.}, ``extended MCA'', in the benchmark 
tables).\\
\indent
The software by Bourke et al.~\cite{BOUR94} fixes the problems of the
original MCA, and it uses a surface template set that exactly agrees to
VESTA, when it is executed in a low resolution (``L''), global 
``disconnect'' mode.
In this particular mode, VESTA will generate surface $N$-cycles up to
length seven only (\textit{cf.}, Fig.s~7 and~8).
In order to be able to better compare the execution times of the codes 
under consideration, all software implementations have been prepared in 
the following ways.\\
\indent
All codes first load a full 3D data set into the computers' memory,
and an initial isovalue has been provided.
Then the start time is taken and the stop watch begins to run.
The codes perform their various tasks and create lists of surface
support points and surface triangles, respectively.
Before all of the generated data are stored into a file that eventually
may be used for, e.g., rendering purposes, the stop watch is halted and
the stop time is taken.
The time, which is the difference time between the stop and start times,
is listed as an average in Tables~2 --~4 for a total of $100$ runs
each.\\
\indent
Note that for this benchmark an Amilo notebook by Fujitsu Siemens
Computers has been used.
Its hardware consists of a Pentium\textregistered
Dual-Core CPU T4200 @ $2.00 GHz$, and $4.00 GB$ RAM.
The disk operating system of this computer is an Ubuntu 10.04 LTS Linux
distribution.

\subsection{Processing of CT-Scan Data of the Stanford Terra-Cotta Bunny}

As a first application, we present in Fig.~26 a VESTA isosurface rendering 
for CT-scan data of the Stanford Terra-Cotta Bunny~\cite{STAN01}.
Note, that this particular rendering uses a flat shading (\textit{cf.},
Ref.~\cite{FOLE93}) for the $150$-valued isosurface, which has been
generated while using VESTA in a high resolution (``H'') ``mixed'' mode
(the latter uses gray-level information to determine locally, whether
a ``disconnect'' or a ``connect'' mode should apply).\\
\indent
In Table 2, all benchmark information is listed.
In particular, all three considered versions of codes yield the same
results with respect to the numbers of surface cycles.
The bunnies' surface is apparently smooth enough, so that $N$-cycles up 
to length six suffice for an adequate result.
Note that the MCA produces the largest amount of redundant information,
i.e., three support points for each created surface triangle,
whereas the original VESTA produces always the minimum of the required
support points.
The marching VESTA avoids in our particular implementation redundant
information only within each given 3D data scanning $3$-cell.\\
\indent
Postprocessing for the removal of redundant support point information
has not been performed by us, since we intended to make only minimal
adaptations to the source code of the MCA implementation by Bourke et
al.~\cite{BOUR94}.
Therefore, this according effort has not been put into the marching VESTA 
implementation either, in order to be able to provide benchmark results
that allow for a better comparison among the codes.
As a result, the here used MCA and marching VESTA codes execute somewhat 
faster than they normally would, if redundant support points were removed.
Note that the original VESTA executes the fastest for this particular
data set (\textit{cf.}, the bold faced ``L"-times in Table~2).
\begin{figure}[!t]
\epsfig{width=8.5cm,figure=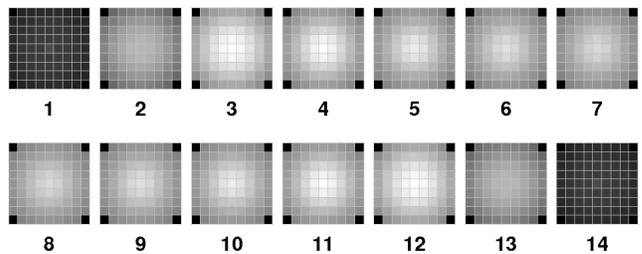}
\vspace{-0.1in} 
\caption{
3D image data set, comprised of an ordered and numbered
stack of $14$ 2D images with $9\times9$ pixels each.
} \label{multib_27}
\end{figure}
\begin{figure}[!b]
\epsfig{width=8.5cm,figure=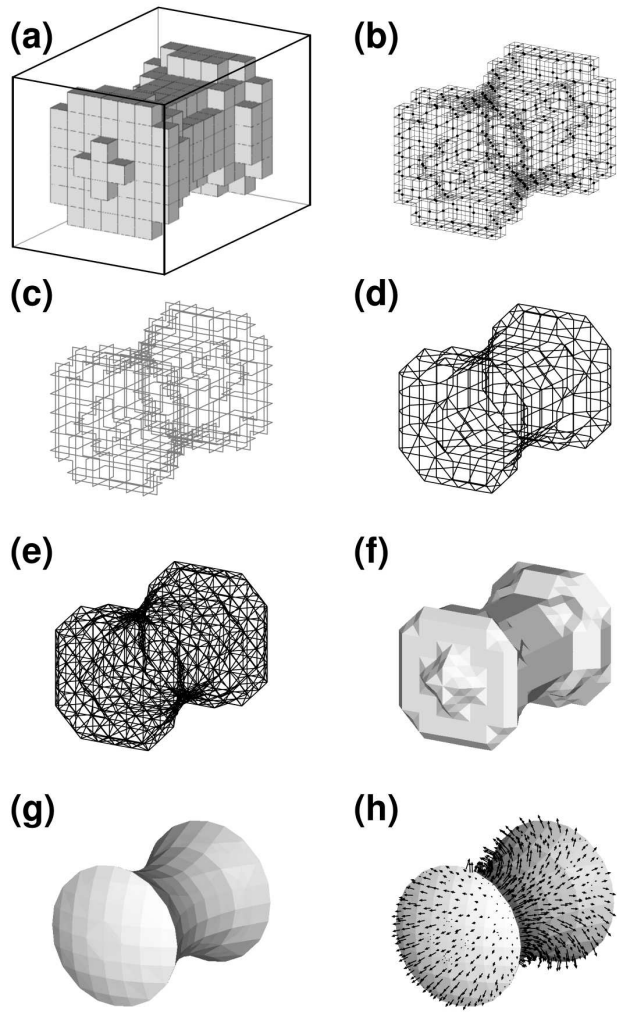}
\vspace{0.0in} 
\caption{
(a) Selected voxels within the $9\times9\times14$ voxel data set as shown 
in Fig.~27; 
(b) initial boundary faces with their centers (black dots)
and voxel face vector pairs (gray lines);
(c) voxel face vector wire frame;
(d) VESTA wire frame for the selected voxels shown in (a); 
(e) wire frame after the breakup of the $N$-cycles into 3-cycles; 
(f) surface rendering for the wire frame shown in (e); 
(g) isosurface rendering using voxel gray-levels for the support
point displacement; 
(h) as in (g) with superimposed surface triangle normal vectors.
} \label{multib_28}
\end{figure}
\begin{figure*}[pt]
\epsfig{width=12.0cm,figure=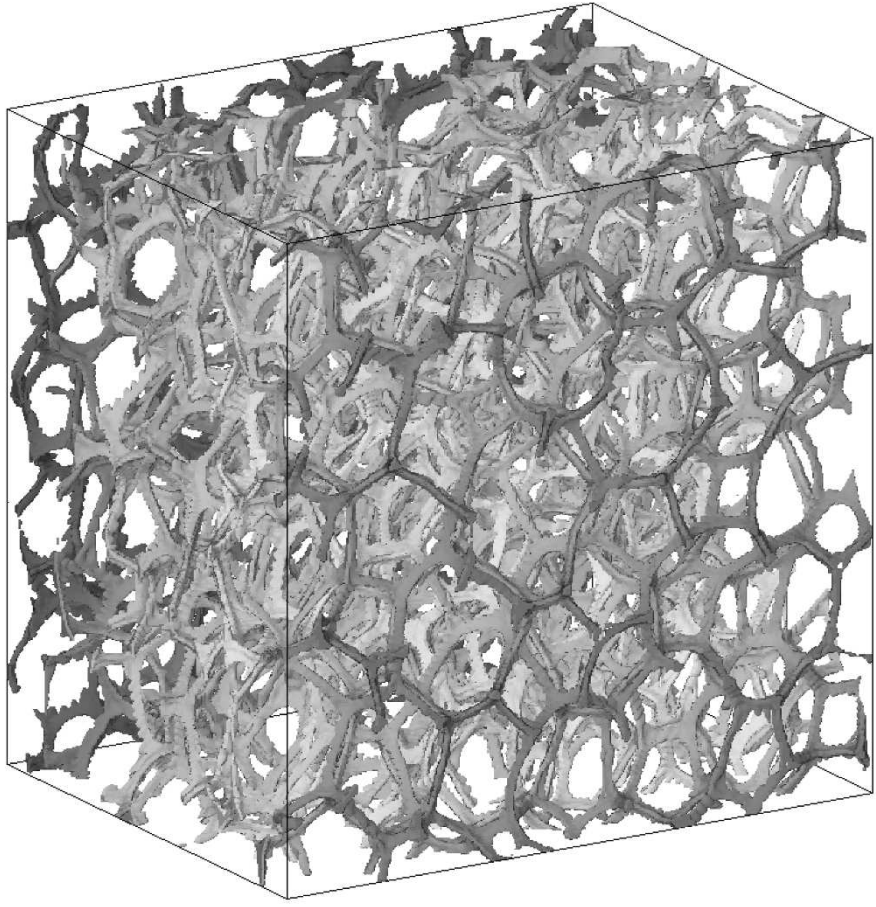}
\vspace{-0.1in} 
\caption{
VESTA isosurface rendering of a foam from 3D image 
data~\cite{SEID03} generated with x-ray microtomography.
} \label{multib_29}
\end{figure*}
\begin{center}
\begin{table*}[pht]
\begin{tabular}{||r||c|c|c||c|c|c||c||}
\hline
\hline
\textbf{Technique}
&\multicolumn{3}{|c||}{original VESTA}
&\multicolumn{3}{|c||}{marching VESTA}
&$\:$extended MCA$\:$\\
\hline
\hline
\textbf{Connectivity$\:$}
&\textbf{Disconnect}&\textbf{Connect}&\textbf{Mixed}
&\textbf{Disconnect}&\textbf{Connect}&\textbf{Mixed}
&\textbf{Disconnect}\\
\hline
\hline
{\boldmath$3$}\textbf{-Cycles}$\:$
&$\: 273,988 \:$
&$\: 272,322 \:$
&$\: 275,741 \:$
&$\: 273,988 \:$
&$\: 272,322 \:$
&$\: 275,741 \:$
&$\: 273,988 \:$\\
\hline
{\boldmath$4$}\textbf{-Cycles}$\:$
&$\: 415,837 \:$
&$\: 419,801 \:$
&$\: 417,010 \:$
&$\: 415,837 \:$
&$\: 419,801 \:$
&$\: 417,010 \:$
&$\: 415,837 \:$\\
\hline
{\boldmath$5$}\textbf{-Cycles}$\:$
&$\: 185,836 \:$
&$\: 185,836 \:$
&$\: 185,066 \:$
&$\: 185,836 \:$
&$\: 185,836 \:$
&$\: 185,066 \:$
&$\: 185,836 \:$\\
\hline
{\boldmath$6$}\textbf{-Cycles}$\:$
&$\: 37,749 \:$
&$\: 40,564 \:$
&$\: 37,804 \:$
&$\: 37,749 \:$
&$\: 40,564 \:$
&$\: 37,804 \:$
&$\: 37,749 \:$\\
\hline
{\boldmath$7$}\textbf{-Cycles}$\:$
&$\: 8,300 \:$
&$\: 4,336 \:$
&$\: 5,589 \:$
&$\: 8,300 \:$
&$\: 4,336 \:$
&$\: 5,589 \:$
&$\: 8,300 \:$\\
\hline
{\boldmath$8$}\textbf{-Cycles}$\:$
&$\: N/A \:$
&$\: N/A \:$
&$\: 1,539 \:$
&$\: N/A \:$
&$\: N/A \:$
&$\: 1,539 \:$
&$\: N/A \:$\\
\hline
{\boldmath$9$}\textbf{-Cycles}$\:$
&$\: N/A \:$
&$\: N/A \:$
&$\: 26 \:$
&$\: N/A \:$
&$\: N/A \:$
&$\: 26 \:$
&$\: N/A \:$\\
\hline
{\boldmath$12$}\textbf{-Cycles}$\:$
&$\: N/A \:$
&$\: N/A \:$
&$\: 0 \:$
&$\: N/A \:$
&$\: N/A \:$
&$\: 0 \:$
&$\: N/A \:$\\
\hline
\textbf{Cycle Sum}$\:$
&$\: 921,710 \:$
&$\: 922,859 \:$
&$\: 922,775 \:$
&$\: 921,710 \:$
&$\: 922,859 \:$
&$\: 922,775 \:$
&$\: 921,710 \:$\\
\hline
\hline
\textbf{L: Points}$\:$
&{\boldmath$\: 932,739 \:$}
&$\: 932,739 \:$
&$\: 932,739 \:$
&{\boldmath$\: 3,699,086 \:$}
&$\: 3,699,086 \:$
&$\: 3,700,651 \:$
&{\boldmath$\: 5,566,998 \:$}\\
\hline
\textbf{L: Triangles}$\:$
&$\: 1,855,666 \:$
&$\: 1,853,368 \:$
&$\: 1,856,666 \:$
&$\: 1,855,666 \:$
&$\: 1,853,368 \:$
&$\: 1,856,666 \:$
&$\: 1,855,666 \:$\\
\hline
\textbf{L: Time [s]}$\:$
&{\boldmath$\: 5.34(1) \:$}
&$\: 5.20(1) \:$
&$\: 5.30(1) \:$
&{\boldmath$\: 4.62(26) \:$}
&$\: 4.66(26) \:$
&$\: 4.58(26) \:$
&{\boldmath$\: 4.71(34) \:$}\\
\hline
\hline
\textbf{H: Points}$\:$
&$\: 1,580,461 \:$
&$\: 1,583,276 \:$
&$\: 1,579,773 \:$
&$\: 4,346,808 \:$
&$\: 4,349,623 \:$
&$\: 4,346,120 \:$
&$\: N/A \:$\\
\hline
\textbf{H: Triangles}$\:$
&$\: 3,151,110 \:$
&$\: 3,154,442 \:$
&$\: 3,147,604 \:$
&$\: 3,151,110 \:$
&$\: 3,154,442 \:$
&$\: 3,147,604 \:$
&$\: N/A \:$\\
\hline
\textbf{H: Time [s]}$\:$
&$\: 5.23(1) \:$
&$\: 5.17(1) \:$
&$\: 5.12(1) \:$
&$\: 5.18(36) \:$
&$\: 5.17(36) \:$
&$\: 5.19(36) \:$
&$\: N/A \:$\\
\hline
\hline
\end{tabular}
\caption{
X-ray micro tomography data benchmark: 
$200$ images with dimensions $256\times256$ = $65,536$ pixels each; 
isovalue equals to $135$; 
number of active voxels: $512,603$, i.e., $3.911\%$. 
}
\label{table_03}
\end{table*}
\end{center}

\subsection{Isosurface Rendering for X-Ray Microtomographic Data}

\noindent
In the field of x-ray microtomography~\cite{STOC08}, like 
tomography~\cite{HERM09}, one creates cross-sections during scans of a 3D object.
These cross-sections are 2D images with pixel sizes in the micrometer range.
The stacking of the 2D images results in a 3D image (data set).
Such a 3D image represents a virtual model of the original object.
Hence, x-ray microtomography provides a way to create a virtual model from an 
object without destroying it.\\
\indent
Let us begin with the sequence of $14$ 2D gray-level images with $9\times9$ 
pixels, which is shown in Fig.~27.
After their stacking, they represent a 3D image.
The pixels here assume gray values in the range from $0$ to $255$.
These bounds represent black and white, respectively.
In the following, we shall apply VESTA to this 3D image in order to render 
an isosurface with isovalue $180$.
First, all voxels that have a gray value larger or equal to $180$ are selected.
The cluster of selected voxels is shown in Fig.~28.a, together with the total
volume of the $9\times9\times14$ voxels.
Note that this initial 3D shape of selected voxels is already enclosed with a 
surface that has no holes at all; 
namely the union of all boundary faces that enclose all active voxels.\\
\indent
Here we shall use gray-level information to automate the (consistent) local 
choosing of the ``disconnect'' and the ``connect'' mode, respectively.
If two voxels that ought to be enclosed have only one single voxel edge 
in common, we determine the (here linearly 
interpolated) gray value of corresponding point of ambiguity.
This gray value is evaluated as the average of the gray-levels of 
the four voxels, which share this common edge.\\
\indent
After application of VESTA, but without support point displacement, 
one obtains -- as intermediate steps -- the 
results as illustrated in Fig.s~28.b -- ~28.d.
The wire frame in Fig.~28.d represents the union of all VESTA $N$-cycles 
where $N=3, ..., 12$. 
In Fig.~28.e, a much denser wire frame is shown, which is the result of 
the break down of the $N$-cycles ($N > 3$) into $3$-cycles while making use 
of additional average points as described in the theoretical section
(\textit{cf.}, Fig.s~9 and 14). 
In a next step, the VESTA $3$-cycles are depicted as solid triangles
(\textit{cf.}, Fig.~28.f). 
Note that the shade of gray of each triangle is determined through the
evaluation of its normal vector.\\
\indent
The preliminary VESTA surface as shown in Fig.~28.f looks somewhat bulky.
However, we should stress that all VESTA surfaces are at this stage of
the processing perfect in the sense, that they are non-degenerate, i.e., they
always enclose a volume that is larger than zero, and they never 
self-intersect and/or overlap each other.
Furthermore, the information of the interior/exterior of the enclosed shapes
is propagated correctly at all times.\\
\indent
In a final step, we generate an isosurface from the VESTA surface.
This is done by displacement of all voxel centers within the bounds defined
by their range vectors, $r$, (\textit{cf.}, Fig.~2) and by subsequent 
reevaluation of the average points, which have been used for the $N$-cycle 
breakup.
In particular, the gray-levels of the two voxels, which define a given
range vector, $r$, are interpolated linearly here (note that other
techniques may be applied~\cite{BOEH07, SALO05}).\\
\indent
The final isosurface representing an isovalue of $180$ for the given 3D image 
data is shown in Fig.~28.g. 
And in Fig.~28.h, this isosurface is superimposed with normal vectors 
of the visible triangles.
Note that this isosurface is quite smooth considering the rather coarse
granularity of the underlying 3D voxel space.
Furthermore, the directions of the normal vectors of the surface elements, 
i.e., triangles, are not limited to the six directions of the normals of the
initial boundary faces.\\
\indent
While using the processing steps as outlined in this subsection, in Fig.~29,
the rendering of a VESTA isosurface with isovalue $135$ is shown for 3D 
x-ray microtomographic image data of a metallic foam~\cite{SEID03}.
In order to better perceive the depth of the data, the rendered surface has 
been color encoded prior to its transformation into shades of gray.
The corresponding benchmark results are presented in Table~3.
Note that the original VESTA needs a little more computing time than the
marching VESTA and the MCA codes, however it generates significantly lesser
surface support points.

\subsection{Freezeout Hypersurface Extraction for Expanding Fireballs}

\noindent
One of the foremost objectives within the field of heavy-ion physics is
the exploration of the equation of state (EOS) of nuclear 
matter (\textit{cf.}, Ref.~\cite{FRIM10} and references therein).
In heavy-ion physics experiments so-called ``fireballs'' are created, which
are very hot and dense zones of nuclear matter.
The theoretical description of these fireballs includes -- but is not 
limited to -- a relativistic hydrodynamic component, in which an EOS has
to be explicitly postulated (\textit{cf.}, e.g., Ref.s~\cite{STRO86, CSER94}).
The resulting space-time evolution of the fireball is driven by the EOS,
and hence, is expected to manifest itself in calculated multi-particle 
production spectra that will eventually be compared to experimental data.\\

\begin{figure}[!t]
\epsfig{width=8.5cm,figure=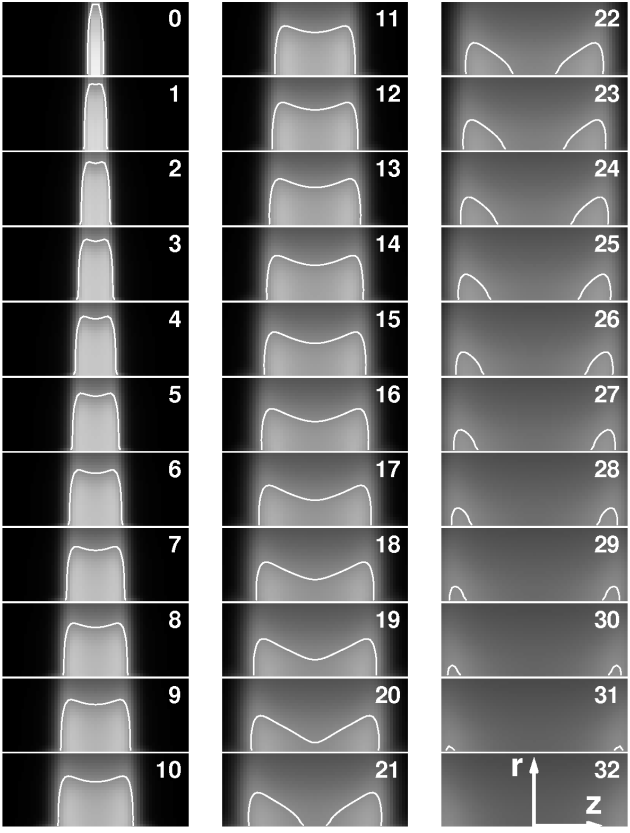}
\vspace{-0.1in} 
\caption{
A temporal sequence of temperature fields as a function of the
longitudinal direction, $z$, and the radius, $r$.
Each frame (except the latter) is superimposed with one or two (white) 
freezeout isocontours with isovalue $T_f = 139 MeV$ (see text). 
} \label{multib_30}
\end{figure}
\begin{figure}[!b]
\epsfig{width=8.5cm,figure=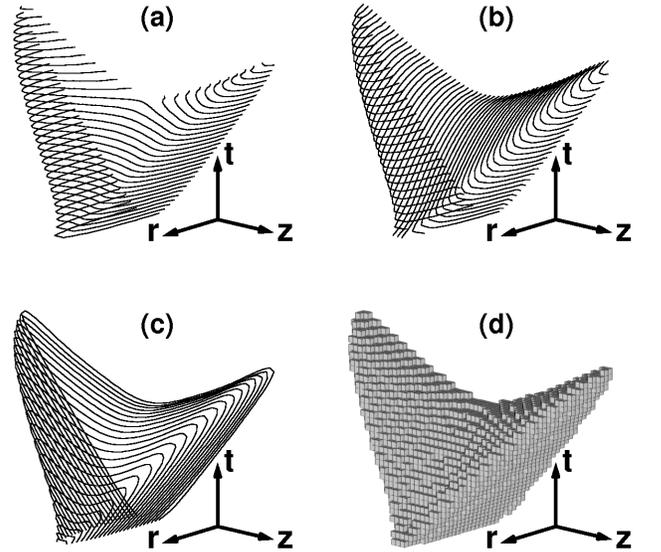}
\vspace{-0.1in} 
\caption{
(a) - (c) Stacked DICONEX freezeout contours, 
(a) of the $z-r$ planes at fixed $t$ values, 
(b) of the $r-t$ planes at fixed $z$ values, and (c) of the $z-t$ planes at 
fixed $r$ values;
note, that each line segment of the contours consists of two
antiparallel vectors, since in 3D both orientations are necessary
when building a surface;
(d) the 3D stack of voxels, which each represent a temperature that is
equal to or higher than $T_f$ (see text). 
} \label{multib_31}
\end{figure}
\begin{figure*}[!t]
\epsfig{width=12.0cm,figure=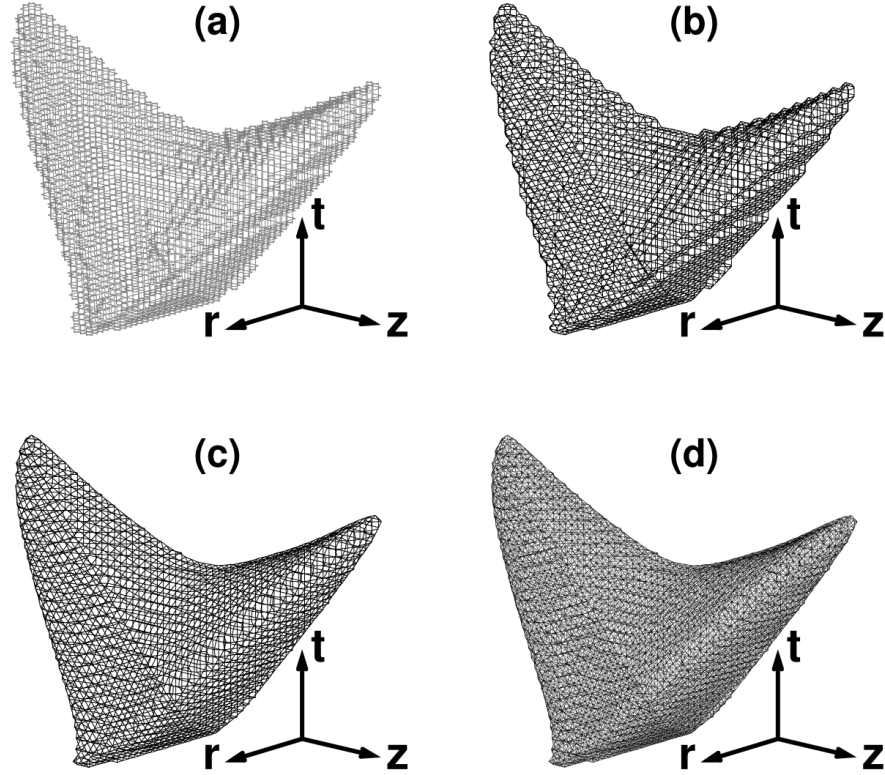}
\vspace{-0.1in} 
\caption{
Various intermediate results of the VESTA processing:
(a) wire frame of voxel face vectors;
(b) VESTA wire frame for the selected voxels shown in Fig.~31.d;
(c) wire frame after support point displacement; 
(d) wire frame after the breakup of the $N$-cycles into 3-cycles (see text).
} \label{multib_32}
\end{figure*}
\begin{figure}[!b]
\epsfig{width=8.5cm,figure=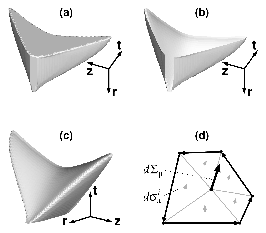}
\vspace{-0.1in} 
\caption{
(a) VESTA surface rendering for the wire frame shown in Fig.~32.d;
(b) as in (a), but after the removal of unphysical triangles;
(c) as in (b), but rotated VESTA freezeout hypersurface with 
$T_f = 139 MeV$ for the 2+1D hydrodynamic simulation data; 
(d) decomposed VESTA $5$-cycle with its corresponding normal vectors (see text).
} \label{multib_33}
\end{figure}

\indent
In some theoretical descriptions (\textit{cf.}, e.g., 
Ref.s~\cite{COOP75, CHEN10}), 
the space-time evolution of the (ultra-relativistic) fluids is in part 
represented through a so-called ``freezeout hypersurface'' (FOHS), which is an 
isosurface within space-time, e.g., with respect to the temperature of the fluid.
The FOHS has to be extracted from the hydrodynamic simulation
data so that further (numerical) calculations can be performed.
Let us consider here the $2+1$ dimensional relativistic hydrodynamic simulation 
code HYLANDER-C~\cite{BRS97, BRS99} that can be used to study radially 
symmetric, so-called ``central'', heavy-ion collisions.\\
\indent
The numerical HYLANDER-C simulations are performed on a cartesian grid in the 
two spatial (2D) variables, $r$ and $z$, and in the temporal (+ 1D) variable $t$.
In particular, $r$ denotes the radius of the system, $z$ the longitudinal 
direction, and $t$ the time, respectively.
A solution of the hydrodynamic equations may be viewed as 3D image data
where each voxel contains continuous values for each of the various physical field
quantities under consideration, such as temperature, $T$, energy density, 
$\epsilon$, components of the fluid velocity $\vec{u}$, etc.
Before we treat 3D simulation data, we shall first take a look at a temporal 
sequence of spatial 2D data.\\
\indent
In Fig.~30, a temporal sequence of $33$ temperature fields is shown 
as a function of the longitudinal direction, $z$, and the radius, $r$.
The time steps remain the same between two successive frames, which are
numbered from $0$ to $32$.
The hydrodynamic grid has originally a much higher resolution than the here 
shown down-sampled 2D images.
Each 2D frame in the figure has a resolution of $61\times24 = 1,464$ pixels,
but eventually we shall process a slightly larger 3D data set (\textit{cf.},
Table~4).
Note that this lower resolution provides sufficient numerical accuracy
for the desired FOHS in this given example. 
Black pixels correspond to a fluid temperature of $0 MeV$, whereas white pixels 
refer to a fluid temperature of $255 MeV$ and/or higher.\\
\indent
For all frames DICONEX isocontours with isovalue (``freezeout temperature'', 
$T_f = $) $139 MeV$ have been extracted (white contours).
The last frame, no. $32$, does not contain any isocontour, 
since all of its pixels have temperatures below $T_f$.
Note that unphysical line segments have been removed where (at least) one end 
point has a radial value that is smaller than zero (\textit{cf.}, 
also Ref.~\cite{BRS09} on the extraction of a FOHS in 2D).\\
\indent
If one stacks up the DICONEX freezeout contours of Fig.~30 in 3D with the 
intent to construct a surface (\textit{cf.}, Fig.~31.a), one requires both 
orientations of the contours.
The latter is a consequence of the fact that in 3D one can look upon a 2D pixel
in two ways, e.g., from ``above'' and from ``below''.
However, if we consider in Fig.~30 the transition from frame no. $20$ to frame
no. $21$, we observe a \textit{correspondence problem}.
Apparently, a single isocontour has to be transformed into two isocontours,
but it is unclear, how it can be accomplished within this approach.\\
\indent
Fig.s~31.b and 31.c offer the two alternate ways to build stacks of isocontours
from the other possible projections within the 3D space under consideration.
Again, unphysical line segments have been removed in the figures where 
(at least) one end point has a radial and/or a temporal value that is smaller 
than zero.
For our particular example, no correspondence problems occur this time, but
it cannot be ensured for different scenarios.
Note that before the invention of VESTA in the year 2002~\cite{BRS03}, FOHS 
construction within HYLANDER-C was initiated while using stacks of isocontours 
from the $r-t$ planes at fixed $z$ values as shown in Fig.~31.b.\\
\begin{center}
\begin{table*}[!ht]
\begin{tabular}{||r||c|c|c||c|c|c||c||}
\hline
\hline
\textbf{Technique}
&\multicolumn{3}{|c||}{original VESTA}
&\multicolumn{3}{|c||}{marching VESTA}
&$\:$extended MCA$\:$\\
\hline
\hline
\textbf{Connectivity$\:$}
&\textbf{Disconnect}&\textbf{Connect}&\textbf{Mixed}
&\textbf{Disconnect}&\textbf{Connect}&\textbf{Mixed}
&\textbf{Disconnect}\\
\hline
\hline
{\boldmath$3$}\textbf{-Cycles}$\:$
&$\: 726 \:$
&$\: 726 \:$
&$\: 726 \:$
&$\: 726 \:$
&$\: 726 \:$
&$\: 726 \:$
&$\: 726 \:$\\
\hline
{\boldmath$4$}\textbf{-Cycles}$\:$
&$\: 1,850 \:$
&$\: 1,850 \:$
&$\: 1,850 \:$
&$\: 1,850 \:$
&$\: 1,850 \:$
&$\: 1,850 \:$
&$\: 1,850 \:$\\
\hline
{\boldmath$5$}\textbf{-Cycles}$\:$
&$\: 532 \:$
&$\: 532 \:$
&$\: 532 \:$
&$\: 532 \:$
&$\: 532 \:$
&$\: 532 \:$
&$\: 532 \:$\\
\hline
{\boldmath$6$}\textbf{-Cycles}$\:$
&$\: 96 \:$
&$\: 96 \:$
&$\: 96 \:$
&$\: 96 \:$
&$\: 96 \:$
&$\: 96 \:$
&$\: 96 \:$\\
\hline
{\boldmath$7$}\textbf{-Cycles}$\:$
&$\: 0 \:$
&$\: 0 \:$
&$\: 0 \:$
&$\: 0 \:$
&$\: 0 \:$
&$\: 0 \:$
&$\: 0 \:$\\
\hline
{\boldmath$8$}\textbf{-Cycles}$\:$
&$\: N/A \:$
&$\: N/A \:$
&$\: 0 \:$
&$\: N/A \:$
&$\: N/A \:$
&$\: 0 \:$
&$\: N/A \:$\\
\hline
{\boldmath$9$}\textbf{-Cycles}$\:$
&$\: N/A \:$
&$\: N/A \:$
&$\: 0 \:$
&$\: N/A \:$
&$\: N/A \:$
&$\: 0 \:$
&$\: N/A \:$\\
\hline
{\boldmath$12$}\textbf{-Cycles}$\:$
&$\: N/A \:$
&$\: N/A \:$
&$\: 0 \:$
&$\: N/A \:$
&$\: N/A \:$
&$\: 0 \:$
&$\: N/A \:$\\
\hline
\textbf{Cycle Sum}$\:$
&$\: 3,204 \:$
&$\: 3,204 \:$
&$\: 3,204 \:$
&$\: 3,204 \:$
&$\: 3,204 \:$
&$\: 3,204 \:$
&$\: 3,204 \:$\\
\hline
\hline
\textbf{L: Points}$\:$
&{\boldmath$\: 3,326 \:$}
&$\: 3,326 \:$
&$\: 3,326 \:$
&{\boldmath$\: 12,814 \:$}
&$\: 12,814 \:$
&$\: 12,814 \:$
&{\boldmath$\: 19,218 \:$}\\
\hline
\textbf{L: Triangles}$\:$
&$\: 6,406 \:$
&$\: 6,406 \:$
&$\: 6,406 \:$
&$\: 6,406 \:$
&$\: 6,406 \:$
&$\: 6,406 \:$
&$\: 6,406 \:$\\
\hline
\textbf{L: Time [s]}$\:$
&{\boldmath$\: 0.124(2) \:$}
&$\: 0.124(1) \:$
&$\: 0.123(1) \:$
&{\boldmath$\: 0.108(1) \:$}
&$\: 0.105(1) \:$
&$\: 0.108(1) \:$
&{\boldmath$\: 0.096(1) \:$}\\
\hline
\hline
\textbf{H: Points}$\:$
&$\: 5,804 \:$
&$\: 5,804 \:$
&$\: 5,804 \:$
&$\: 15,292 \:$
&$\: 15,292 \:$
&$\: 15,292 \:$
&$\: N/A \:$\\
\hline
\textbf{H: Triangles}$\:$
&$\: 11,362 \:$
&$\: 11,362 \:$
&$\: 11,362 \:$
&$\: 11,362 \:$
&$\: 11,362 \:$
&$\: 11,362 \:$
&$\: N/A \:$\\
\hline
\textbf{H: Time [s]}$\:$
&$\: 0.122(1) \:$
&$\: 0.124(1) \:$
&$\: 0.123(3) \:$
&$\: 0.109(2) \:$
&$\: 0.109(1) \:$
&$\: 0.109(1) \:$
&$\: N/A \:$\\
\hline
\hline
\end{tabular}
\caption{
Relativistic hydrodynamic simulation data benchmark: 
$33$ images with dimensions $241\times81$ = $19,521$ pixels each; 
isovalue equals to $139$; 
number of active voxels: $8,000$, i.e., $1.242\%$.
}
\label{table_04}
\end{table*}
\end{center}
\vspace*{-1.1cm}
\indent
In order to avoid any correspondence problems within 3D, the $2+1$D 
relativistic hydrodynamic simulation code HYLANDER-C starts nowadays from the 
consideration of the full 3D stack of voxels, which each represent a 
temperature that is equal to or higher than $T_f$ (\textit{cf.}, Fig.~31.d).
The application of VESTA (\textit{cf.}, Fig.~32), while using average points
for the breakup of the $N$-cycles into $3$-cycles as explained above, yields the 
isosurface for the isovalue $T_f = 139 MeV$, which is depicted in 
Fig.s~33.b and~33.c.
During the process of FOHS construction, all unphysical triangles have been 
discarded, i.e., those, which have (at least) one corner with a radial and/or a 
temporal value that is smaller than zero.\\
\indent
Note that the union of all isocontour edges, which are shown in 
Fig.s~31.a -- 31.c, is also given through the set of triangular VESTA 
surface mesh edges as shown in Fig.~32.c (without the unphysical ones) 
and vice versa.
The gray shading of each triangle in Fig.s~33.a --~33.c is determined from 
the knowledge of the normal vector of each of the triangles of the FOHS.
Benchmark comparisons between the various VESTA and MCA codes are provided
in Table~4.
Although the extended MCA is here the fastest, it also creates the most
redundant information, i.e., $19,218$ support points instead of only 
$3,326$.\\
\indent
Since the number of normal vectors can be quite large for a generated FOHS, it 
may be desirable to reduce this number for the speed up of subsequent, further 
calculations.
In Fig.~33.d, a single decomposed VESTA $5$-cycle is shown.
To each triangle, its normal vector (gray), $d\sigma_{\mu}^{i}$ 
($i = 1, ..., 5$), has been attached to the average (or ``center of mass'') 
point of each the three triangle corner points.
Note that the orientations of these normal vectors have been inherited from 
the initial orientation of the VESTA $5$-cycle.\\
\indent
In order to prevent further averaging of other field quantities, which are 
potentially present within the initial 3D voxel data set, and because we
intend to speed up subsequent calculations, it is adequate to construct the 
total normal vector, $d\Sigma_{\mu}$, as the sum of the triangle 
normals, $d\sigma_{\mu}^{i}$
\vspace*{-0.1cm}
\begin{equation}
	d\Sigma_{\mu} = \sum_{i = 1}^N d\sigma_{\mu}^{i}
\end{equation}

\noindent
where $d\Sigma_{\mu}$ is associated with the average (or ``center of mass'') 
point of each support point of the VESTA $N$-cycle under consideration.
Note that in the case of VESTA $3$-cycles, 
$d\Sigma_{\mu} \equiv d\sigma_{\mu}^{1}$.
In this case, $d\Sigma_{\mu}$ is associated with the center of mass point of 
the corresponding single triangle.
For further considerations of normal vector construction, we would like to 
refer the reader to Ref.~\cite{CHEN10}.\\

\section{Summary}

\noindent
In summary, all the details that are necessary for a successful implementation of 
the volume-enclosing surface extraction algorithm, which has been named VESTA, have
been explained here for the very first time. 
VESTA surfaces are always perfect in the sense, that they are non-degenerate, i.e.,
they always fully enclose a volume that is larger than zero, and they never 
self-intersect and/or overlap each other
(i.e., prior to a possible move of the initial boundary face centers within the
bounds of the corresponding range vectors).\\
\indent
We would like to stress that holes can never occur within a VESTA surface.
Hence, VESTA surfaces do not require any kind of repair since potential
ambiguities are correctly resolved.
Furthermore, the information of the interior/exterior of the enclosed shapes
is propagated also correctly at all times.
In particular, the VESTA surfaces may be viewed as the 3D analog of the 
DICONEX contours in 2D.
Note that VESTA could be extended to adaptive and/or unstructured 3D
grids as well.
Furthermore, it is fully compatible with its 4D counterpart, 
STEVE~\cite{BRS04}, which is capable of the processing of time-varying voxel
data. 
In fact, the ideas which have been presented in this paper can be
applied to $n$-dimensional spaces~\cite{PAT11}.\\
\indent
The fact that VESTA surface cycles are confined to $2\times2\times2$ voxel
neighborhoods is a result, but not a prerequisite as it is the case in 
Refs.~\cite{LORE87, BLOO94, HO05}.
Unlike the mesh generators presented in Ref.s~\cite{LORE87, HO05}, VESTA is not
template-based.
Instead, VESTA uses the single building block, which is shown in Fig.~2.b.
VESTA will find all of the required surface segments, which are determined 
with a fast and 100\% robust construction technique.\\
\indent
We have demonstrated the usage of VESTA for several rather diverse 
applications, 
namely for the creation of isosurfaces based on 3D image data in the fields
computer tomography and x-ray microtomography, respectively, as well as for 
the creation of a freezeout hypersurface from a set of 3D numerical simulation
data in the field of relativistic heavy-ion physics.
Various benchmarks have shown that VESTA can easily compete with the
Marching Cubes algorithm, e.g., as far as computing speed is concerned.
In addition, VESTA can produce six different types of surface outputs.
\section{Acknowledgements}

\noindent
This work has been supported by the Department of Energy under contract
W-7405-ENG-36.
I am indebted to Dr. Dan Strottman for the extended support after my 
return to Germany.

\bigskip

\bibliographystyle{elsarticle-harv}

\end{document}